\documentclass{aa}
\usepackage{graphicx}
\usepackage{txfonts}
\newcommand{\Si}[5]{\mbox{$#1\,^#2{\rm #3}^{{\rm #4}}_{\rm #5}$}}

\newcommand{\SH}{$S_{\!\!\rm H}$}

\newcommand{\eps}[1]{\log\varepsilon_{\rm #1}}

\begin{document}
\title{\textbf{Statistical equilibrium of silicon in the solar atmosphere}
\thanks{Table 1 is only available in electronic form at
http://www.edpsciences.org}}
\author{J.R. Shi\inst{1,2}  \and T. Gehren\inst{2} \and K. Butler\inst{2} \and L.I. Mashonkina\inst{2,3}
 \and G. Zhao\inst{1,2}} \offprints{J.R. Shi,\ \
 \email{sjr@bao.ac.cn}}

\institute {National Astronomical Observatories, Chinese Academy of Sciences,
Beijing 100012, P.R. China  \and Institut f\"ur Astronomie und Astrophysik der
Universit\"at M\"unchen, Scheinerstr. 1, D-81679 M\"unchen, Germany  \and
Institute of Astronomy, Russian Academy of Sciences, Pyatnitskaya Str. 48,
Moscow, Russia 109017 \\}

\date{Received / Accepted }

\abstract
{}
{The statistical equilibrium of neutral and ionised silicon in the
solar photosphere is investigated. Line formation is discussed and
the solar silicon abundance determined.}
{High-resolution solar spectra were used to determine solar $\log
gf\varepsilon_{\rm Si}$ values by comparison with Si line synthesis
based on LTE and NLTE level populations. The results will be used in
a forthcoming paper for differential abundance analyses of
metal-poor stars. A detailed analysis of silicon line spectra leads
to setting up realistic model atoms, which are exposed to
interactions in plane-parallel solar atmospheric models. The
resulting departure coefficients are entered into a line-by-line
analysis of the visible and near-infrared solar silicon spectrum.}
{The statistical equilibrium of \ion{Si}{i} turns out to depend
marginally on bound-free interaction processes, both radiative and
collisional. Bound-bound interaction processes do not play a
significant role either, except for hydrogen collisions, which have
to be chosen adequately for fitting the cores of the near-infrared
lines. Except for some near-infrared lines, the NLTE influence on
the abundances is weak.}
{Taking the deviations from LTE in silicon into account, it is
possible to calculate the ionisation equilibrium from neutral and
ionised lines. The solar abundance based on the experimental
$f$-values of Garz corrected for the Becker et al.'s measurement is
$7.52 \pm 0.05$. Combined with an extended line sample with selected
NIST $f$-values, the solar abundance is $7.52 \pm 0.06$, with a
nearly perfect ionisation equilibrium of
$\Delta\log\varepsilon_\odot(\ion{Si}{ii}/\ion{Si}{i}) = -0.01$. }
{}

 \maketitle

\section{Introduction}
Neutral silicon gives rise to only a very few and predominantly weak
spectral lines in the visible. For that reason it is sometimes
regarded as unspectacular and consequently ignored in many abundance
analyses. There are at least three important facts that put Si back
into perspective:
\begin{itemize}
    \item Silicon is commonly attributed to the $\alpha$ elements that are assumed
    to be produced in massive type II supernovae. In this case the Galactic
    chemical evolution of Si should differ significantly from that of iron
    resulting in logarithmic overabundance ratios [Si/Fe] $> 0$.
    \item Silicon belongs to the most abundant metals. It competes with Mg and Fe, which also
    have similar ionisation energies. It is therefore important as an electron
    donor in relatively cool \emph{turnoff} stellar atmospheres, where elements
    such as H, C, N, O and Ne with even higher abundances cannot contribute.
    \item Due to its condensation behaviour, Si is found in many interstellar
    and interplanetary dust particles and meteorites. Therefore, Si is the
    reference element for abundance determinations from solar system meteorites.
    This is different from the stellar abundance reference, hydrogen.
\end{itemize}

All three points cannot hide the problem every astronomer encounters
when trying to determine the Si abundance in cool turnoff stars.
Even the line spectrum observed in metal-rich stars is poor, which
may be part of the reason why the laboratory work is also far from
complete, and some terms and line transitions have not yet been
identified unambiguously. On the other hand, spectra of hotter stars
such as those of types B or O display a multitude of spectral lines
of up to three consecutive ionisation stages, \ion{Si}{ii},
\ion{Si}{iii}, and \ion{Si}{iv}.

Neutral silicon has only a single strong line in the blue part of
the spectrum, i.e. the \Si{3p}{1}{S}{}{0} -- \Si{4s}{1}{P}{o}{1}
3905 \AA\ line. That line appears in a spectral range that is not
easy to observe because
\begin{itemize}
    \item cool stars have their maximum flux in the red,
    \item spectrograph coatings are optimized for the visual, and
    \item CCD detectors have a lower QDE in the blue.
\end{itemize}

\begin{figure*}
\begin{center}
\resizebox{15cm}{!}{\includegraphics{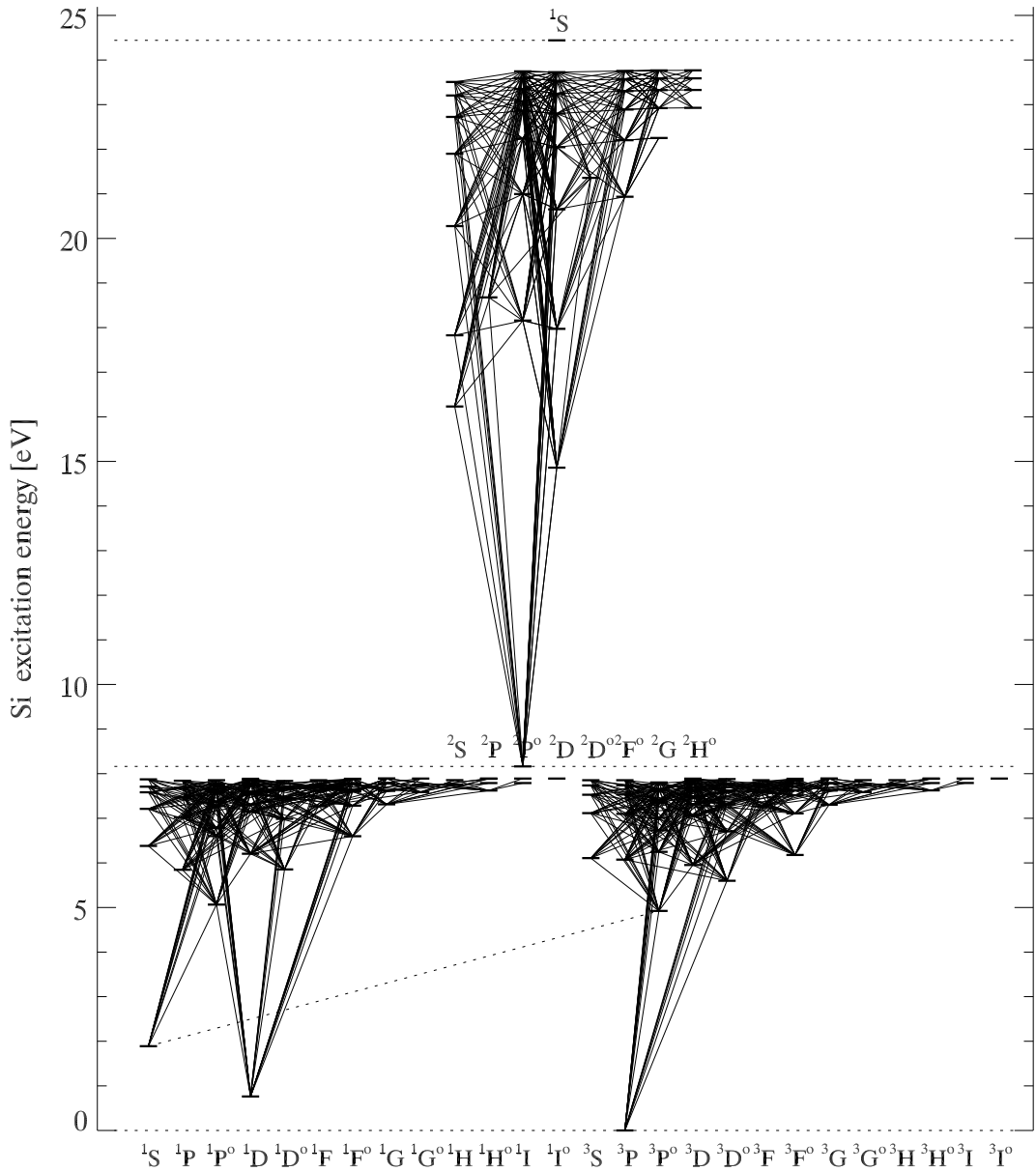}} \caption[]{Term
energy diagram of the silicon model atom used in this analysis with
\ion{Si}{ii} (top) and \ion{Si}{i} (bottom). \ion{Si}{ii} quartets
are neglected. Allowed transitions are continuous, the forbidden
\ion{Si}{i} 4103 \AA\ line is dotted.}
\end{center}
\label{fig1}
\end{figure*}
Additionally, the line is blended by a number of other metal lines,
and line synthesis requires a careful representation of the blend
components, in particular in metal-rich stellar spectra. A second
important line is the intercombination line, \Si{3p}{1}{S}{}{0} --
\Si{4s}{3}{P}{o}{1} at 4103 \AA. Different from other allowed
transitions in the visible and the near red, this line is relatively
strong (comparable to the \ion{Mg}{i} line at 4571 \AA), and at
typical S/N of stellar spectra the detection becomes difficult at
metallicities [Fe/H] $<-$2.5.

\begin{figure*}
\resizebox{\textwidth}{!}{\includegraphics{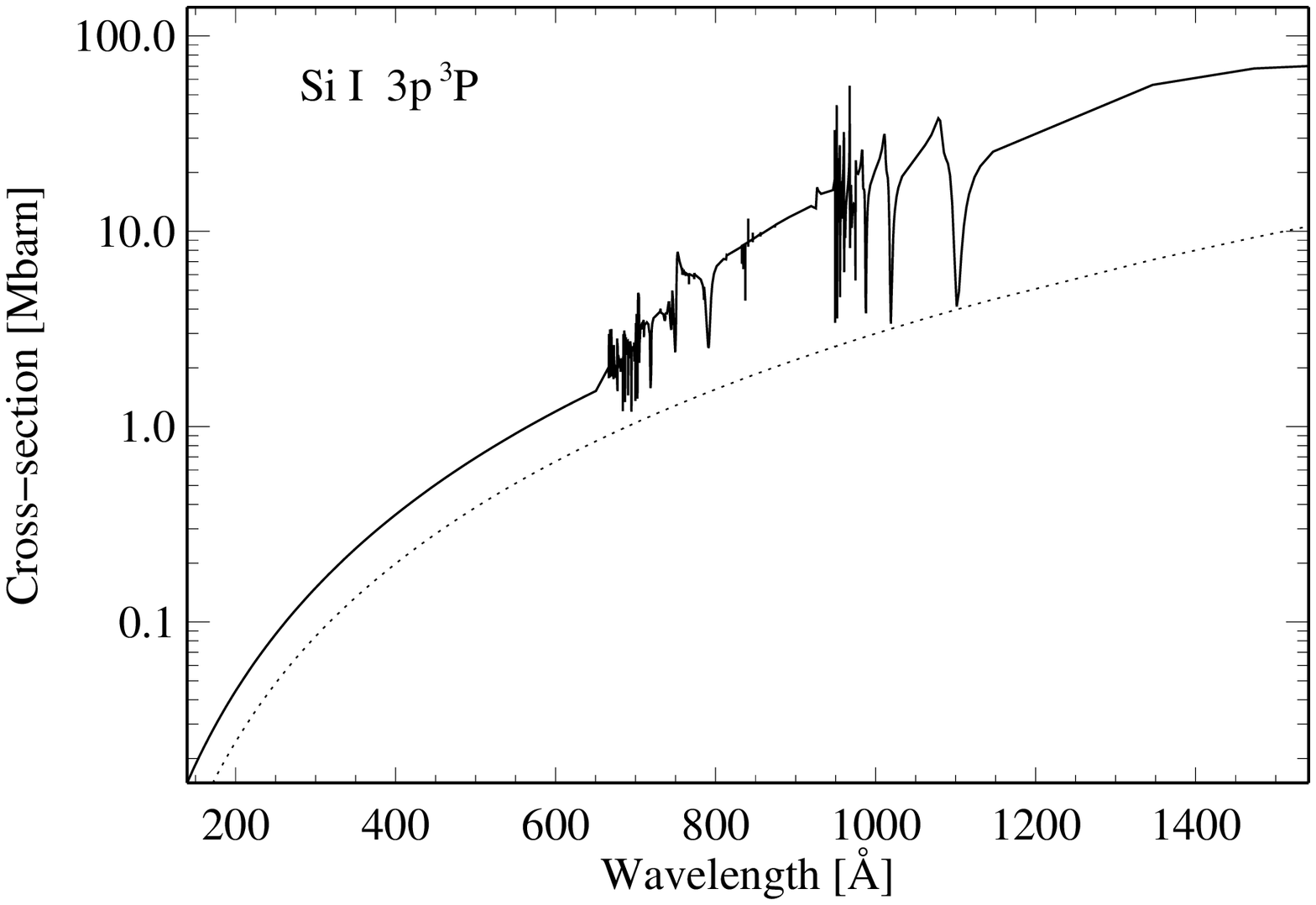}
\includegraphics{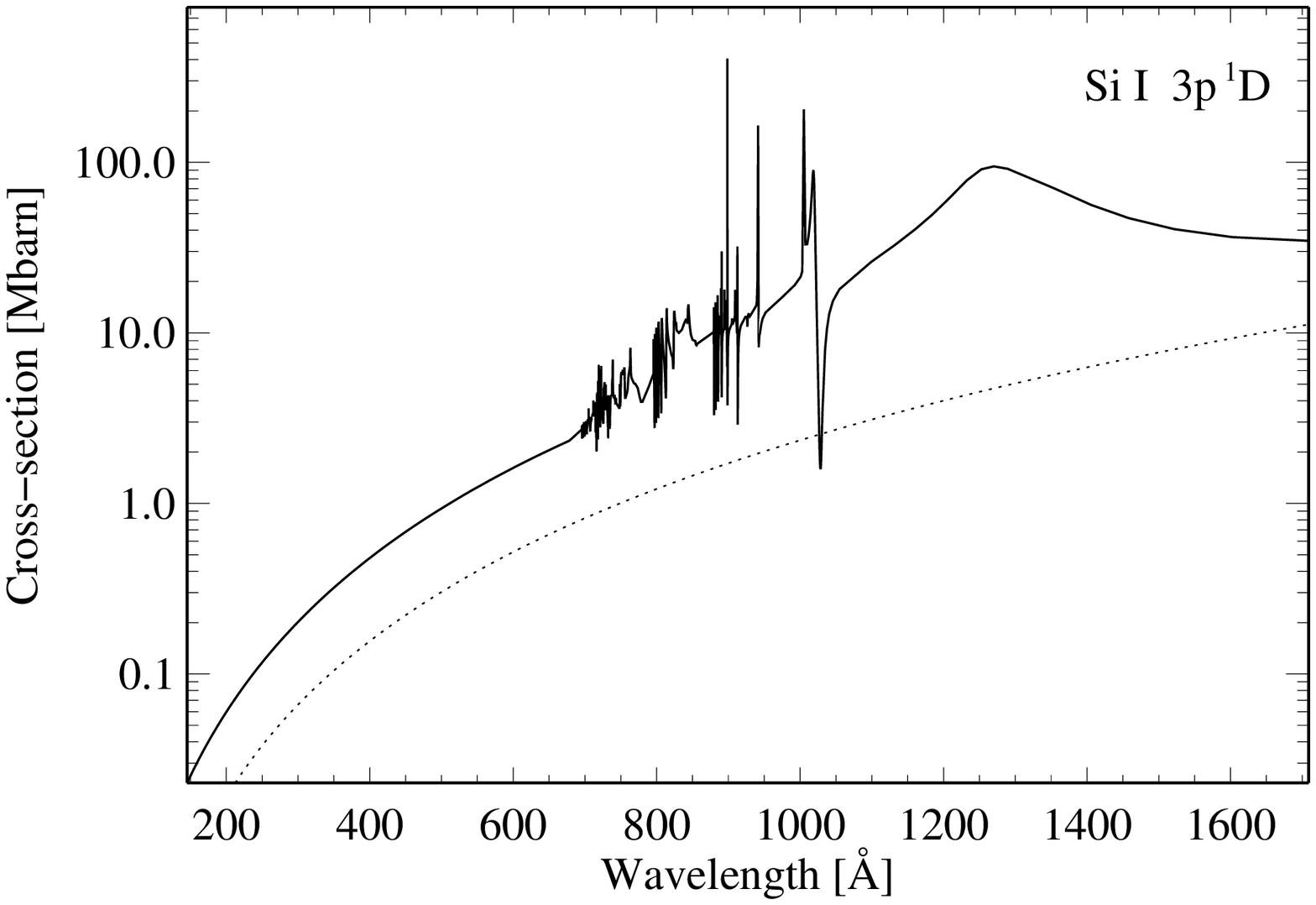}
\includegraphics{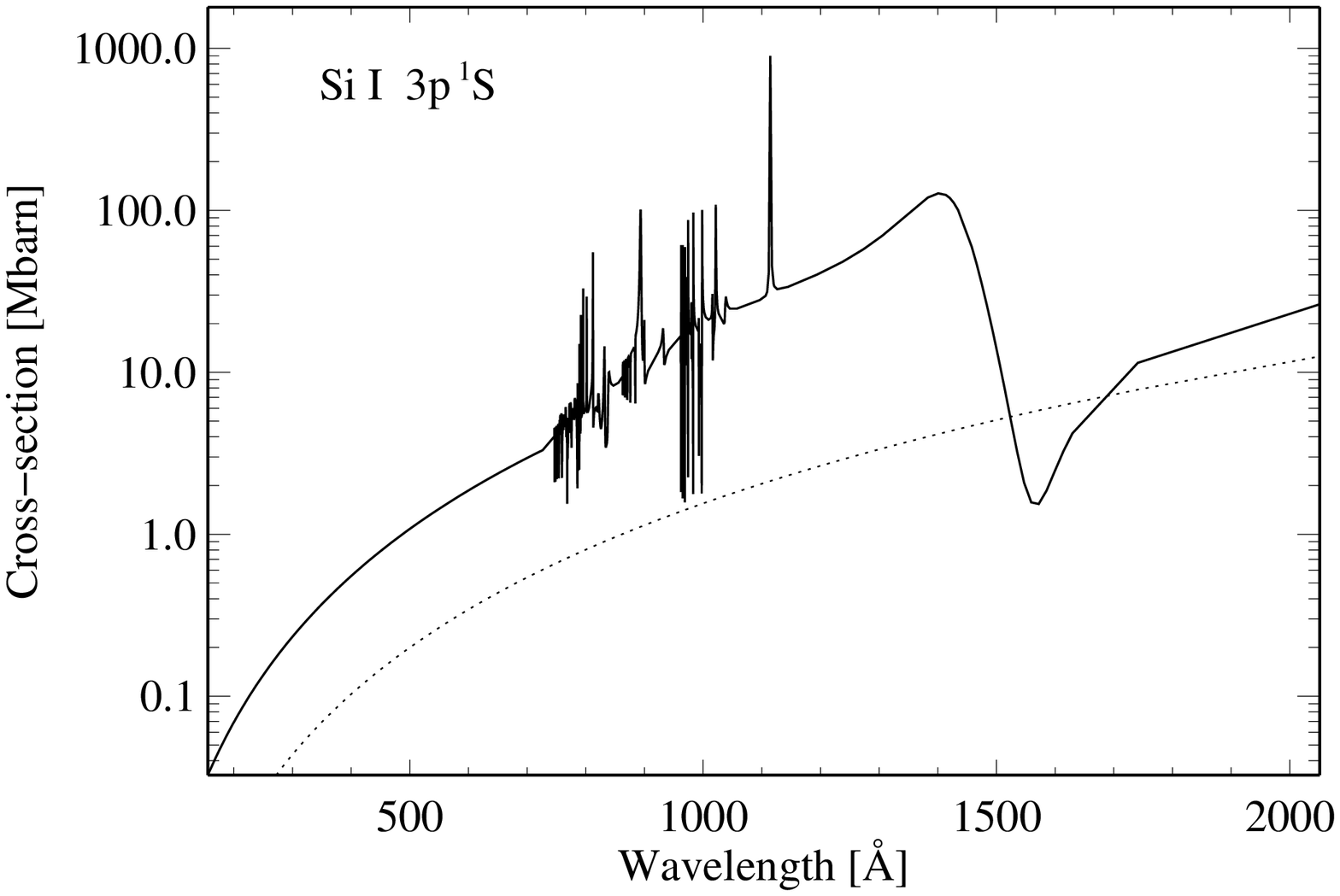}}
\caption[]{Photoionisation cross-sections of the metastable terms of
the \ion{Si}{i} model atom as given in the TOPbase. All three
cross-sections are significantly larger than the corresponding
hydrogenic approximations (dotted curves). The absorption edge is
located at the right boundary of the individual frames.}
\label{fig2}
\end{figure*}
The situation becomes increasingly worse in \emph{metal-poor stars},
even more so if the stars are at the hot end of the turnoff with
effective temperatures around 6200 to 6500 K. Here, \ion{Si}{i} as a
typical minority ion becomes nearly fully ionised leaving
practically no neutral silicon atoms in any of the excited levels.
Since the triplet lines arising from the \Si{3p}{3}{P}{}{} ground
state are all in the satellite UV, they cannot be observed from the
ground. The few near-infrared lines are absorbed from levels above 5
eV and also become invisible in the spectra of metal-poor turnoff
stars.

Whereas the solar spectrum still provides enough absorption lines
for a thorough abundances analysis, this is not the case for hot
metal-poor stars. Even the 4103 \AA\ intercombination line cannot be
used there for an abundance analysis, and silicon abundances of such
stars will have to rely on a single line, $\lambda 3905$ \AA. This
makes an LTE abundance analysis particularly prone to systematic
errors due to atomic interaction processes that tend towards a net
over- or underpopulation of the ground state. As we shall see below,
the \Si{3p}{3}{P}{}{0,1,2} levels and the metastable
\Si{3p}{1}{D}{}{2} and \Si{3p}{1}{S}{}{0} levels are strongly
depopulated by photoionisation, and a NLTE analysis of the line
formation problem is therefore unavoidable.

As mentioned above, silicon spectra of cool stars were not
investigated often. Only one NLTE analysis has appeared to date.
Wedemeyer (\cite{WE01}) reported the influence of deviations from
LTE upon the Si abundances in the Sun and in Vega with no reference
to line formation in metal-poor stars. A number of LTE analyses,
however, cover the whole range of metal abundances, from solar to
the most extreme metal-poor halo stars. Among them are the classical
analysis of solar silicon lines by Holweger (\cite{HH73}) and the
more recent work of Asplund (\cite{AS00}), who introduces 3D line
formation based on a solar hydrodynamical model.

The introduction of the atomic model used is the topic of Section 2.
In Section 3 two atmospheric models of the Sun are presented.
Section 4 shows the results of the NLTE calculations obtained with
different settings of the remaining free parameters. In Section 5
spectral line synthesis under both LTE and NLTE assumption is
applied to determine the photospheric solar Si abundance and the
\ion{Si}{i}/\ion{Si}{ii} ionisation equilibrium. The results are
discussed in the final section.

\section{Atomic model of silicon}
As usual, the \emph{level system} of silicon was extracted from the
NIST data bank\footnote{http://www.physics.nist.gov/}. Silicon in
cool stars is fully represented by the first three ionisation
stages, where for the purpose of NLTE calculations \ion{Si}{iii}
reduces to the ground state, \Si{3s^2}{1}{S}{}{}. The configuration
of the neutral atom includes 4 electrons, with that of the lowest
energy being $3s^2\,3p^2$. This gives rise to singlets and triplets
with a different outer electron replacing $3p$. A single quintet
term is found near 4 eV, which is due to a $3s\,3p^3$ configuration,
and which does not couple to any other singlet or triplet term with
known line strength, except for the \Si{3p}{3}{P}{}{} --
\Si{3p^3}{5}{S}{o}{}. This transition is so weak that the quintet
term can safely be neglected. A similar configuration exists in
ionised silicon, where the quartet term \Si{3s\,3p^2}{4}{P}{}{} at
5.3 eV is ignored. Our silicon atomic system therefore consists only
of the singlet/triplet system of \ion{Si}{i}, the doublet system of
\ion{Si}{ii}, and the singlet ground state of \ion{Si}{iii}. It is
shown in Fig. \ref{fig1} as an energy level diagram with all the
terms included here. To keep the number of lines and levels
manageable, the level system was reduced to terms implying a thermal
differential distribution of the corresponding level populations.
Thus, a total of 132 terms of \ion{Si}{i}, 41 terms of \ion{Si}{ii}
plus the \ion{Si}{iii} ground state give our approach a realistic
background. As with other elements, completeness is lost at high
excitation energies. This holds in particular for the \ion{Si}{ii}
ion, where the last eV is not represented. However, since these
terms are more than 15 eV from the \ion{Si}{ii} ground state, the
population of the levels is extremely low.

\emph{Line transitions} are based on the work of the Opacity Project
(\cite{SE94}), in particular the calculations of Nahar \& Pradhan
(\cite{NP93}). The total number of lines included is 786 for
\ion{Si}{i} and 182 for \ion{Si}{ii}. For NLTE transfer calculations
simple Gauss profiles were used with 9 wavelength points each.

Fortunately, \emph{bound-free transition} cross-sections with
complex structure are available from calculations at the
TOPbase\footnote{http://vizier.u-strasbg.fr/topbase/topbase.html}
(see also Nahar \& Pradhan \cite{NP93}). The cross-sections of the
lowest three terms, \Si{3p}{3}{P}{}{}, \Si{3p}{1}{D}{}{} and
\Si{3p}{1}{S}{}{} are an order of magnitude greater than those of
the next terms, which is very similar to the configuration of Mg
(Zhao et al. \cite{ZH98}). It tends to decouple the metastable terms
efficiently from the excited ones.

\emph{Background opacities} are calculated with an opacity sampling
code based on the hydrogen lines, the line lists made available by
Kurucz (\cite{KU92}), and on the important bound-free cross-sections
of hydrogen and the most abundant metals. The background opacities
are sampled on a \emph{random grid} of between 5000 and 10000
wavelengths, to which are added the wavelengths of the line
profiles. The final NLTE line formation program thus samples roughly
14000 wavelengths. In the UV, where most of the background opacity
is expected, the wavelength intervals of the random grid are between
1 and 4 \AA. We emphasise that the choice of this frequency grid
affects only the \emph{bound-free} radiative interaction rates.
Earlier experiments with a finer frequency mesh of up to 40000
randomly sampled frequencies for \ion{Fe}{i} (Gehren et al.
\cite{GE01}) showed that the influence of a very dense frequency
sampling is less than marginal. This is similar to the sampling
situation for OS model atmospheres, where changes in temperature
corrections drop below a few K when the number of frequencies is
increased to values beyond 10000 (Grupp \cite{GR04}, Fig. 8).
Background opacities for \emph{bound-bound} transitions are sampled
at the corresponding line frequencies.

\begin{figure}
\resizebox{\columnwidth}{!}{\includegraphics{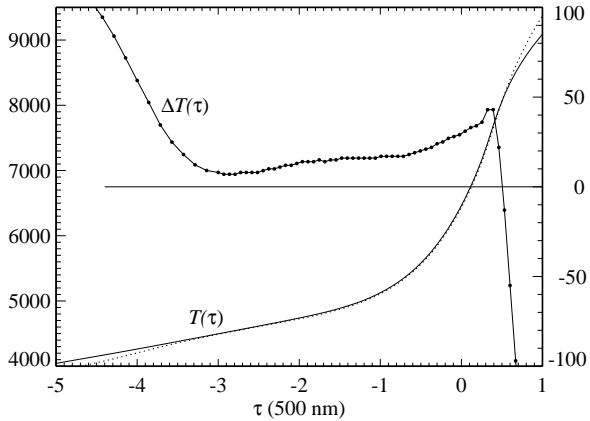}}
\caption[]{Temperature stratifications of ODF (dotted) and OS (solid
line) solar models (left axis). Difference between the two
stratifications, $\Delta T(\tau) = T(\tau) {\rm OS}-T(\tau) {\rm
ODF}$ (right axis).} \label{fig3}
\end{figure}
For our Si NLTE calculations we take into account \emph{inelastic
collisions} with electrons and hydrogen atoms for both excitation
and ionisation. The formulae of van Regemorter (\cite{VR62}) and
Allen ({\cite{AL73}) are used to describe the excitation of allowed
and forbidden transitions by electron collisions, respectively.
Ionisation cross-sections for electron collisions are calculated
applying the formula of Seaton (\cite{SE62}). Drawin's (\cite{DR68},
\cite{DR69}) formula as described by Steenbock \& Holweger
(\cite{SH84}) is used to calculate neutral hydrogen collisions, with
a similar formula for bound-free hydrogen collisions. To allow for
some empirical correction to the Drawin approximations, a scaling
factor $S_{\!\rm H}$ is applied to the formula in our calculations.

\section{Solar model atmospheres}
Solar atmosphere models are supposed to give the basic background
for the NLTE line formation program as well as to the spectrum
synthesis. This investigation assumes a standard 1D model atmosphere
in radiative-convective equilibrium, where convection is treated in
a mixing-length approach. Since there exist two different opacity
approximations based either on opacity distribution functions (ODF)
or on opacity sampling (OS), it seems worthwhile extending our
research to the differences of the silicon level populations between
the two atmospheric models.

For this purpose both solar models are considered. The ODF model is
our old MAFAGS standard atmosphere, nearly identical to that of
Kurucz (\cite{KU92}), but downscaled from Kurucz' value of $\eps{Fe}
= 7.67$ to a solar Fe abundance of $\eps{Fe} = 7.51$. It 
was used in all previous NLTE calculations (e.g. Gehren et al.
\cite{GE01}). The OS model is based on a similar model atmosphere
code of Grupp (\cite{GR04}), but with convection treated according
to Canuto \& Mazzitelli (\cite{CM92}), and with opacities sampled
from continuous opacities and from an extended line list of Kurucz
\& Bell (\cite{KB95}) at 86000 randomly chosen frequency points. The
two types of atmospheres lead to slightly different temperature
stratifications, as can be seen in Fig. \ref{fig3}. The two models
differ in temperature outside $\log\tau \simeq 0.3$ with a
moderately increasing amplitude. Note that the differences outside
$\log\tau = -3$ do not affect the observed Si spectrum, whereas
inside $\log\tau = 0.3$ the gap between the temperature
stratifications is the result of different versions of the
mixing-length theory (see Grupp, 2004, for a more detailed
comparison). The latter marginally affects the formation of the
Balmer lines, but does not influence Si line formation very much.

\section{NLTE line formation results}

\begin{figure}
\resizebox{\columnwidth}{!}{\includegraphics{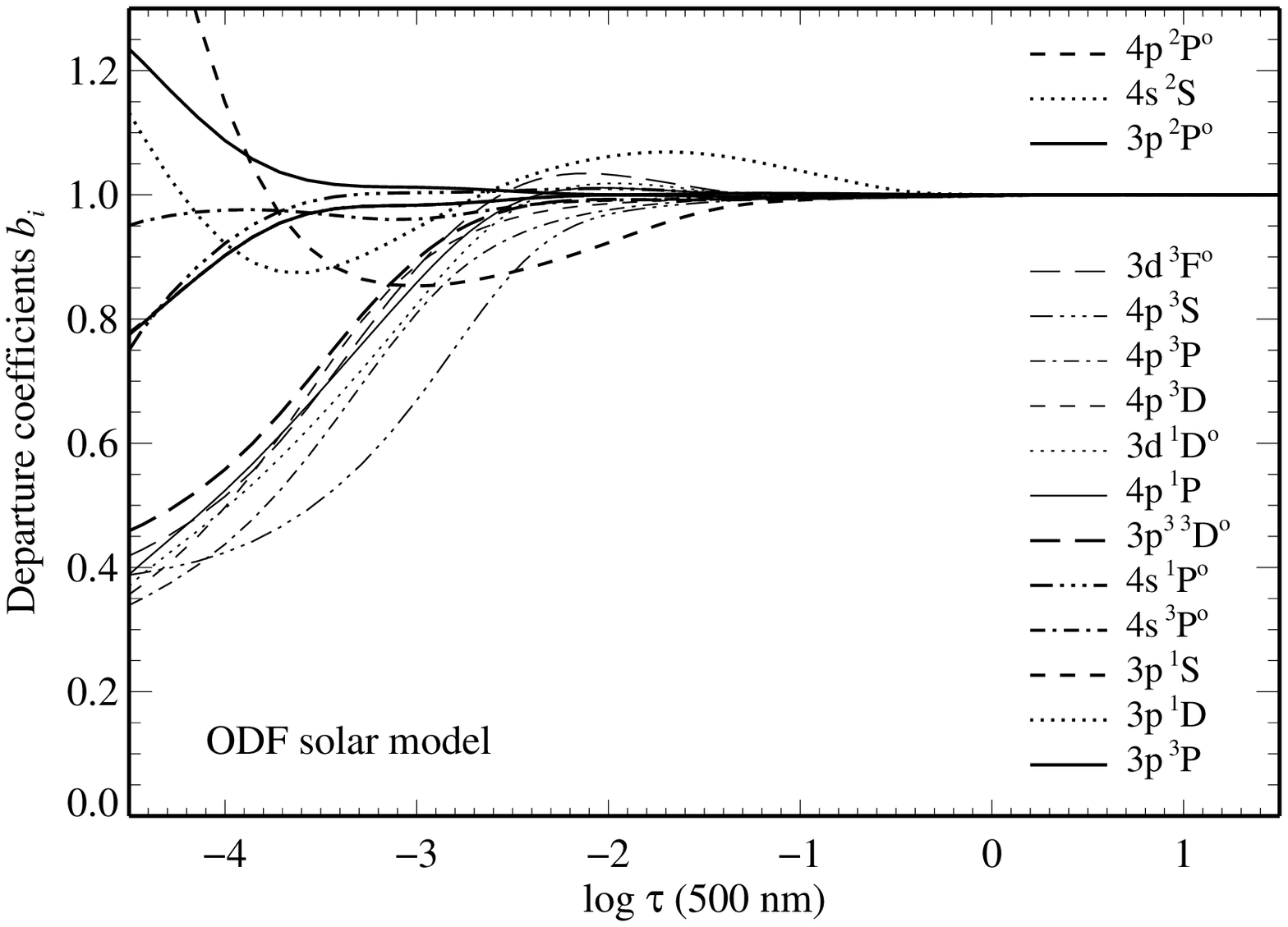}}\\[-5mm]
\resizebox{\columnwidth}{!}{\includegraphics{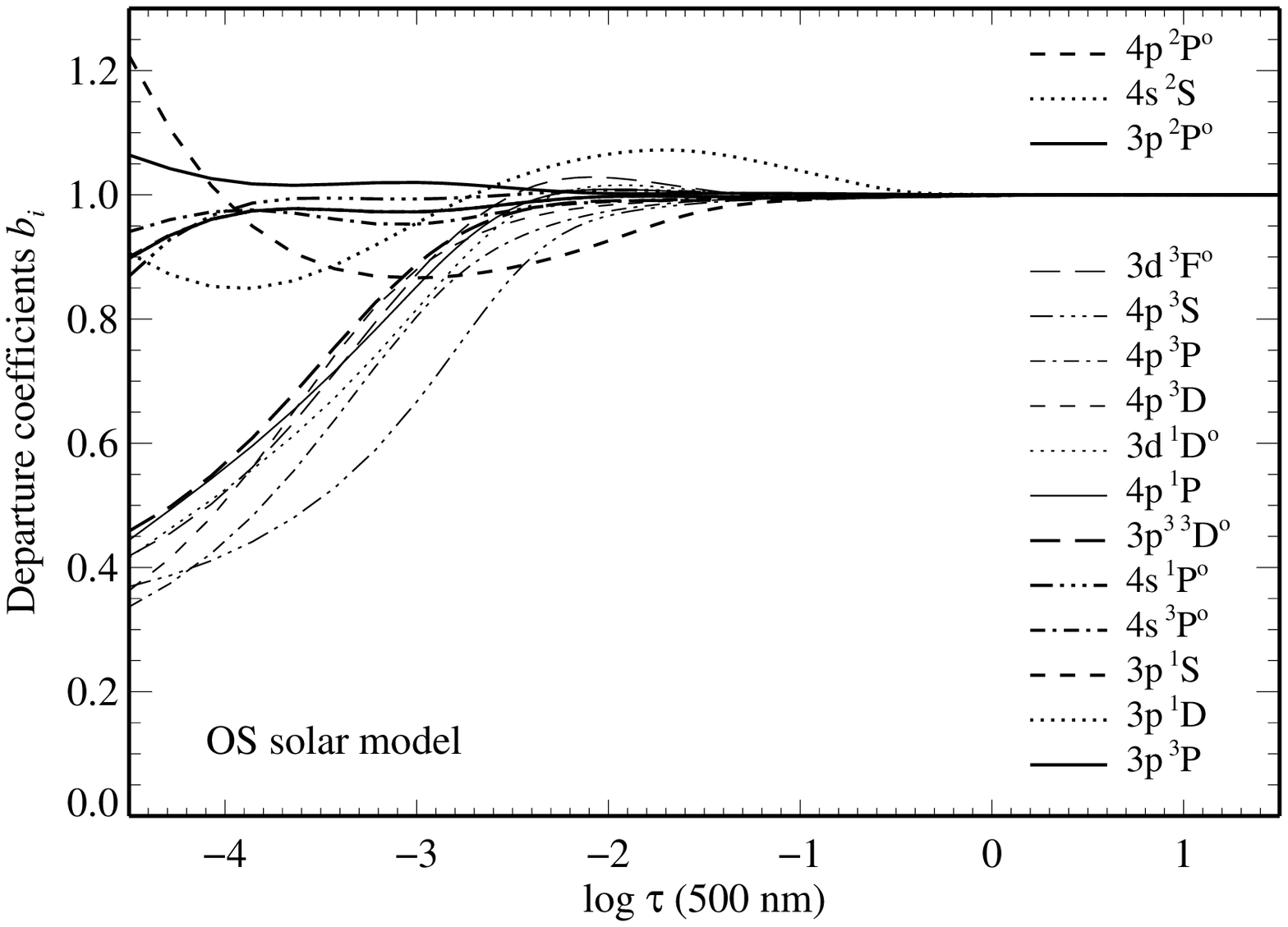}}
\caption[]{Departure coefficients $b i = N i/N i^{\rm LTE}$ for the
final parametrization of the Si atomic model (\SH\ = 0.1) calculated
for an ODF (top) and for an OS (bottom) solar model atmosphere as
function of the standard optical depth. Note the similarity of the
results for both atmospheric models. The doublet terms
\Si{4s}{2}{S}{}{}, \Si{4p}{2}{P}{o}{} belong to \ion{Si}{ii}.}
\label{fig4}
\end{figure}
The Si departure coefficients for the more important terms of
\ion{Si}{i} and \ion{Si}{ii} shown for our final selection of the
hydrogen collision parameter (\SH\ = 0.1) in Fig. \ref{fig4} are
surprising in that they follow LTE out into the middle photosphere.
This has not been encountered in any of the other neutral ions
analysed so far. It is the result of an unusually large energy gap
between the ground state, \Si{3p}{3}{P}{}{}, and the two metastable
terms \Si{3p}{1}{D}{}{}, and \Si{3p}{1}{S}{}{}, on one side, and the
first excited levels of neutral silicon on the other. These gaps are
$\sim$ 5, 4, and 3 eV, respectively, and they shift all lines
emerging from those levels into the UV, most of them below 2000 \AA,
where the \ion{Al}{i} absorption edge is suppressed the radiation
field by an order of magnitude. As a result, all these
strong\footnote{according to their population and $f$-value} lines
have only weak radiative rates due to a very low integrated mean
radiation field. There is a marginal similarity with the \ion{Mg}{i}
ion (Zhao et al. \cite{ZH98}), however, with the difference that the
more important low-excitation transitions in neutral magnesium are
well above 2000 \AA\ and thus considerably stronger. This is also
the case for \ion{Fe}{i} which shares ionisation energy and solar
abundance with both Mg and Si but has a much more complex line
spectrum with strong lines in the visible.

Therefore, the interaction of the three most populated \ion{Si}{i}
levels with the \ion{Si}{ii} ion is completely based on bound-free
processes, where photoionisation and ionisation by electron
collisions compete in strength, with a small net pumping to the
\ion{Si}{ii} ground state. As would be expected, the
\Si{3p}{1}{D}{}{} and \Si{3p}{1}{S}{}{} terms follow
\Si{3p}{3}{P}{}{} tightly. In the absence of stronger radiative
rates the excited terms are determined by collisions in their
respective regions of line formation.

\begin{figure}
\resizebox{\columnwidth}{!}{\includegraphics{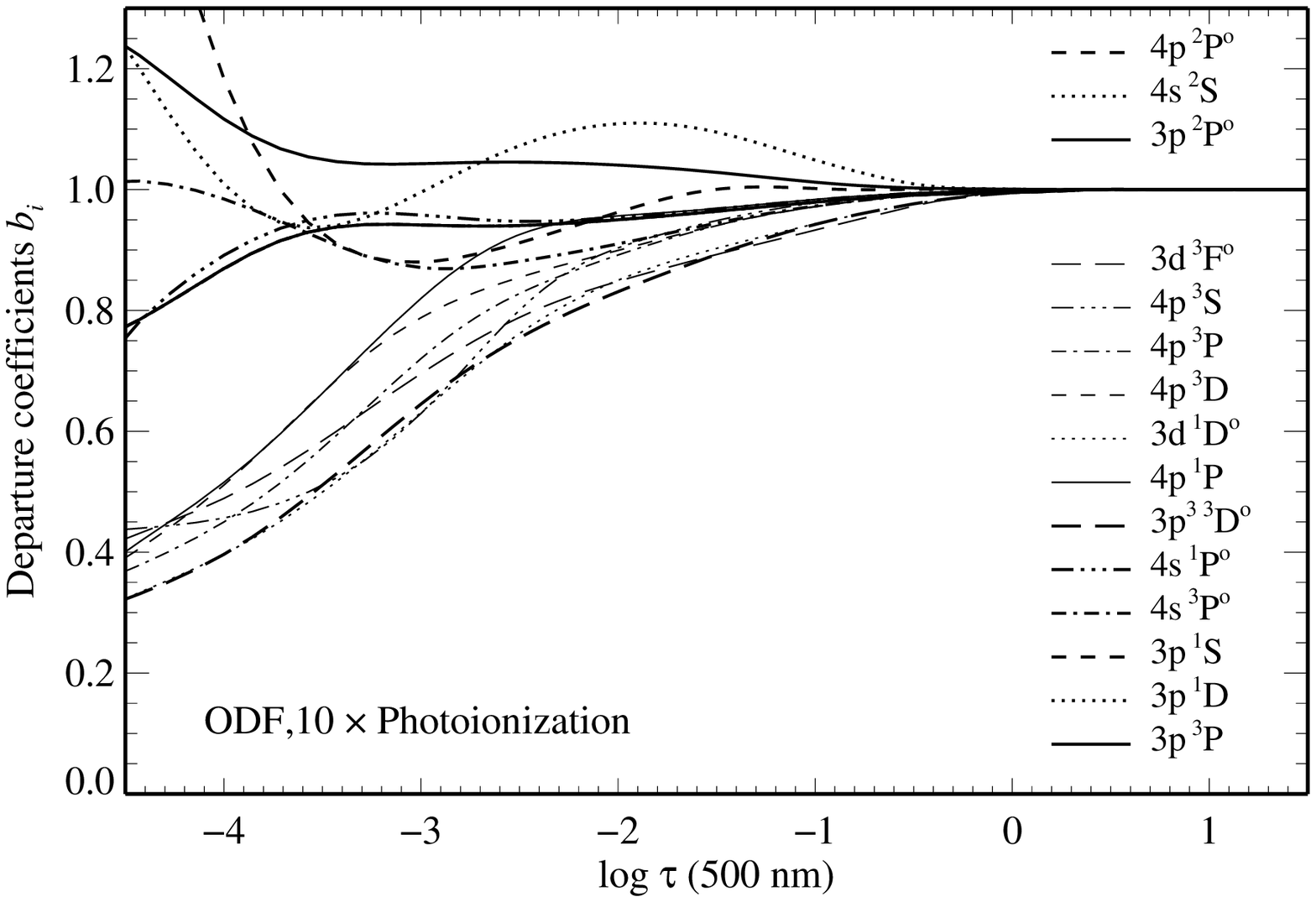}}\\[-5mm]
\resizebox{\columnwidth}{!}{\includegraphics{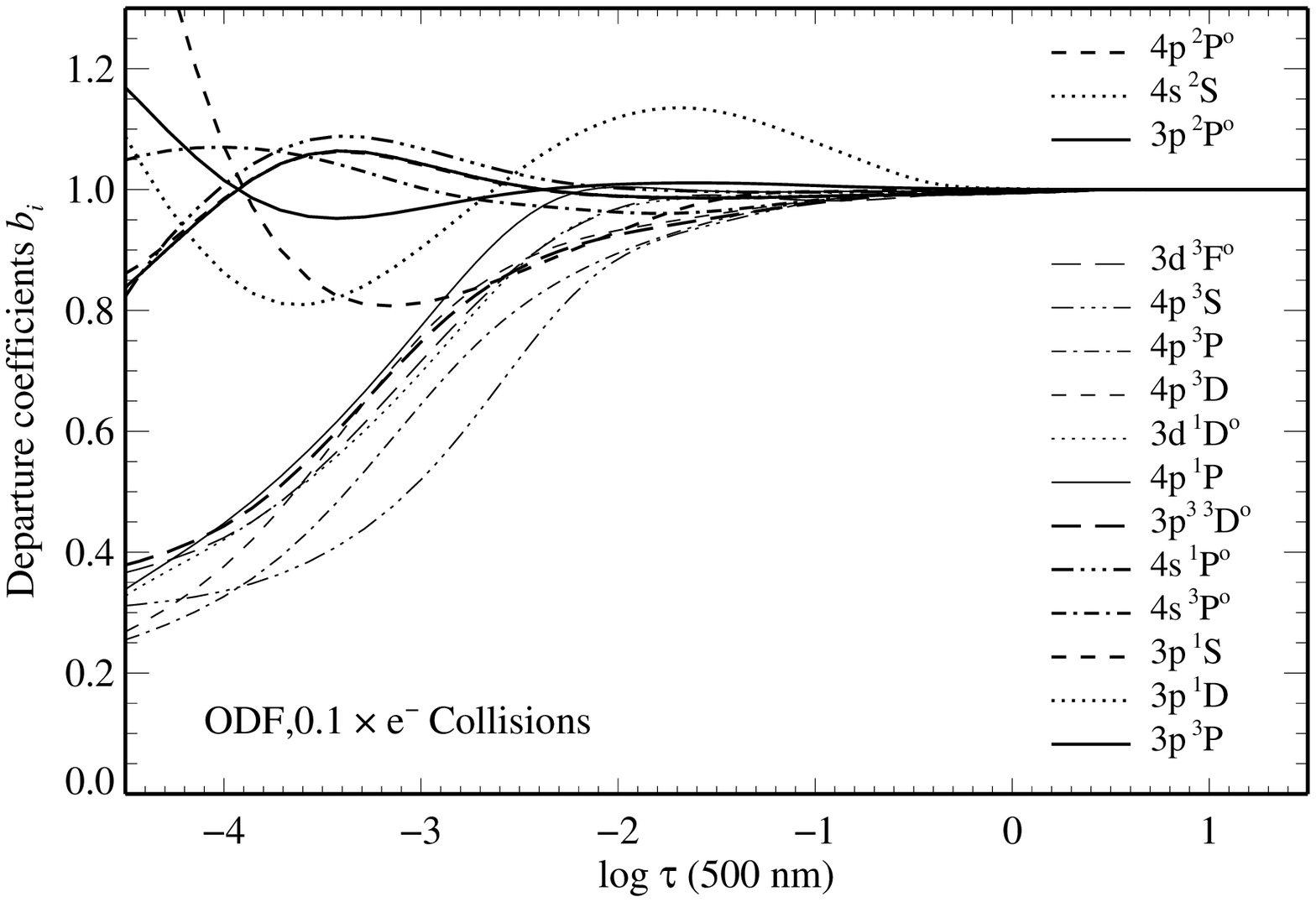}}
\caption[]{Departure coefficients for the final atomic model in the
ODF solar atmosphere. All parameters are the same as in Fig.
\ref{fig4}, except that the photoionisation rates were scaled by a
factor of 10 (top), and the electronic collision rates divided by a
factor of 10 (bottom).} \label{fig5}
\end{figure}
Test calculations applying increased photoionisation or decreased
electronic collisions reveal that \ion{Si}{i} is relatively stable
against a variation of atomic data. Although a strongly increased
photoionisation (see Fig. \ref{fig5}, top) starts to depopulate the
levels already in deeper layers, this does not affect neutral Si
line formation very much, because most of the lines observed in the
visible and near infrared solar spectrum are weak and thus formed in
deeper layers. The only strong line at 3905 \AA\ retains a
thermalised source function out to the upper photosphere. The
ionisation itself changes only by a very small fraction. The bottom
of Fig. \ref{fig5} demonstrates a variation of the electron
collision rates by a factor of 0.1. Similar to the increased
photoionisation, though less pronounced in its influence on the
departure coefficients, the reduction of the electronic collision
efficiency also leads only to a marginal de-thermalisation, that
does not affect neutral silicon line formation very much. Only the
near-infrared lines respond to collisions (see below).

While the \ion{Si}{ii} \Si{3p}{2}{P}{o}{} ground state is a true
mirror of the \ion{Si}{i} \Si{3p}{3}{P}{}{} ground state, the
excited levels of \ion{Si}{ii} lead to absorption lines coming from
much deeper layers in the atmosphere. This holds in particular for
the highly excited \Si{4s}{2}{S}{}{} -- \Si{4p}{2}{P}{o}{}
transition with lines at 6347.095 and 6371.360 \AA, respectively.
Although absorbed at an excitation energy of more than 8 eV, the
lines are clearly visible in the spectra of solar-type stars and may
therefore serve as a test for the ionisation and \ion{Si}{ii}
excitation equilibrium in the Sun.

A variation of the hydrogen collision scaling factor \SH\ between 0
and 1 allows us to introduce similar changes as those resulting from
electron collisions. Due to the large number of hydrogen atoms, the
neutral hydrogen collision rates can be significantly stronger than
the electron collision rates. Yet, in the range of \SH\ parameters
considered here, they do not lead to significant changes in the line
profiles in the visible where, generally, abundance differences for
the two extremes (\SH = 0 and 1) stay within 0.03 \ldots 0.04 dex.
This situation changes significantly, whenever the \ion{Si}{i}
near-infrared lines are analysed. As is evident from Table 1 most of
these are absorbed from either \Si{4s}{3}{P}{o}{},
\Si{4p}{1}{P}{}{}, or \Si{4p}{3}{D}{}{}. Since all the excited terms
are only loosely coupled to each other and to other high-excitation
terms, their departure coefficients tend to diverge from each other
at $\log\tau \simeq -2$, i.e. where the \emph{cores} of the
relatively strong near-infrared lines form. This is the reason for a
slightly increased abundance trend going with the hydrogen collision
scaling factor. For lines beyond 1 $\mu$m the corresponding
abundance variations can reach more than 0.1 dex, and their cores
suggest the choice of \SH\ = 0.1.

All calculations were carried out with a revised version of the
DETAIL program (Butler \& Giddings \cite{BG85}) using accelerated
lambda iteration following the extremely efficient method described
by Rybicki \& Hummer (\cite{RH91}, \cite{RH92}).

\begin{figure*}
\resizebox{\textwidth}{!}{\includegraphics{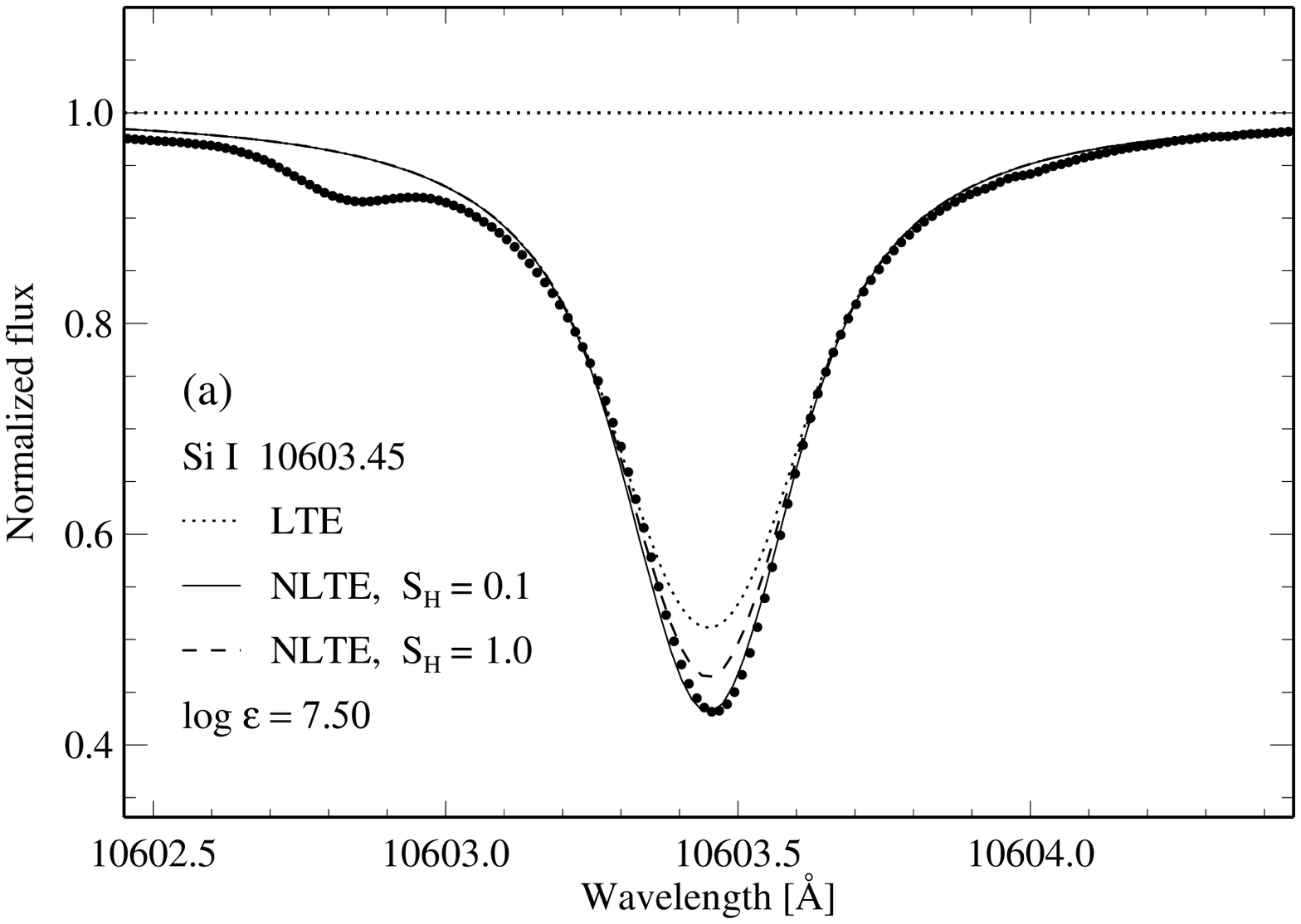}\includegraphics{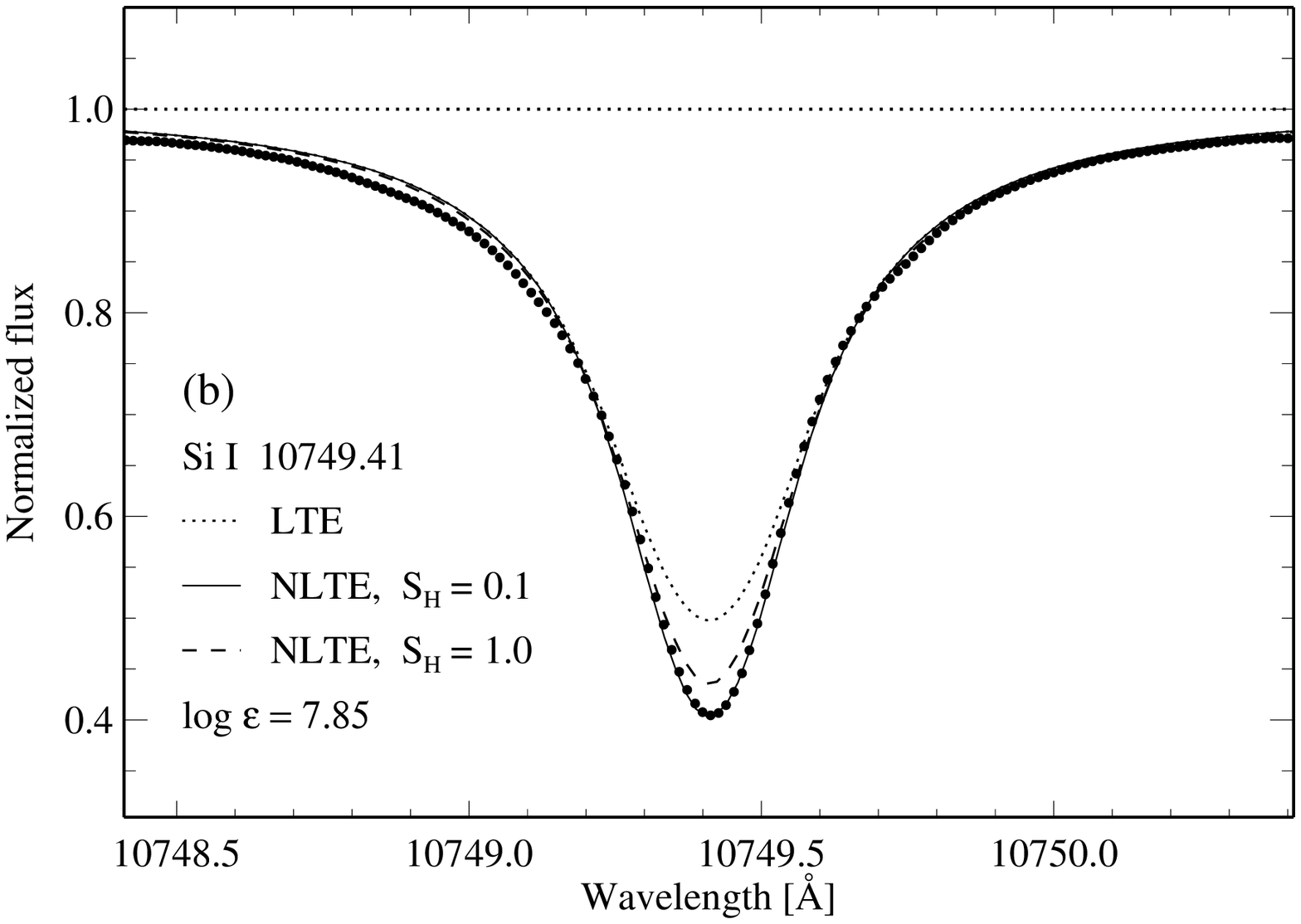}}\\[-0.3cm]
\resizebox{\textwidth}{!}{\includegraphics{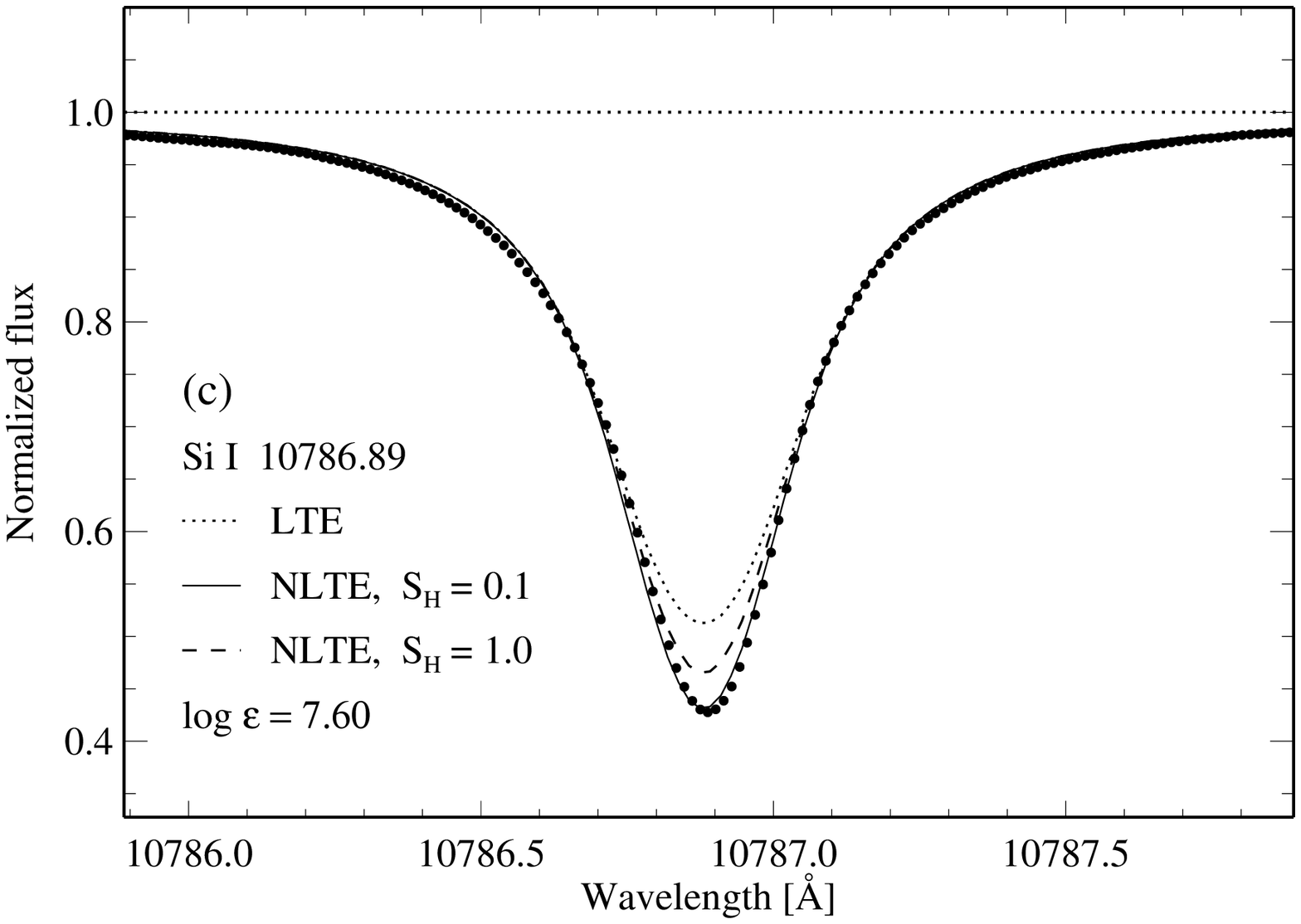}\includegraphics{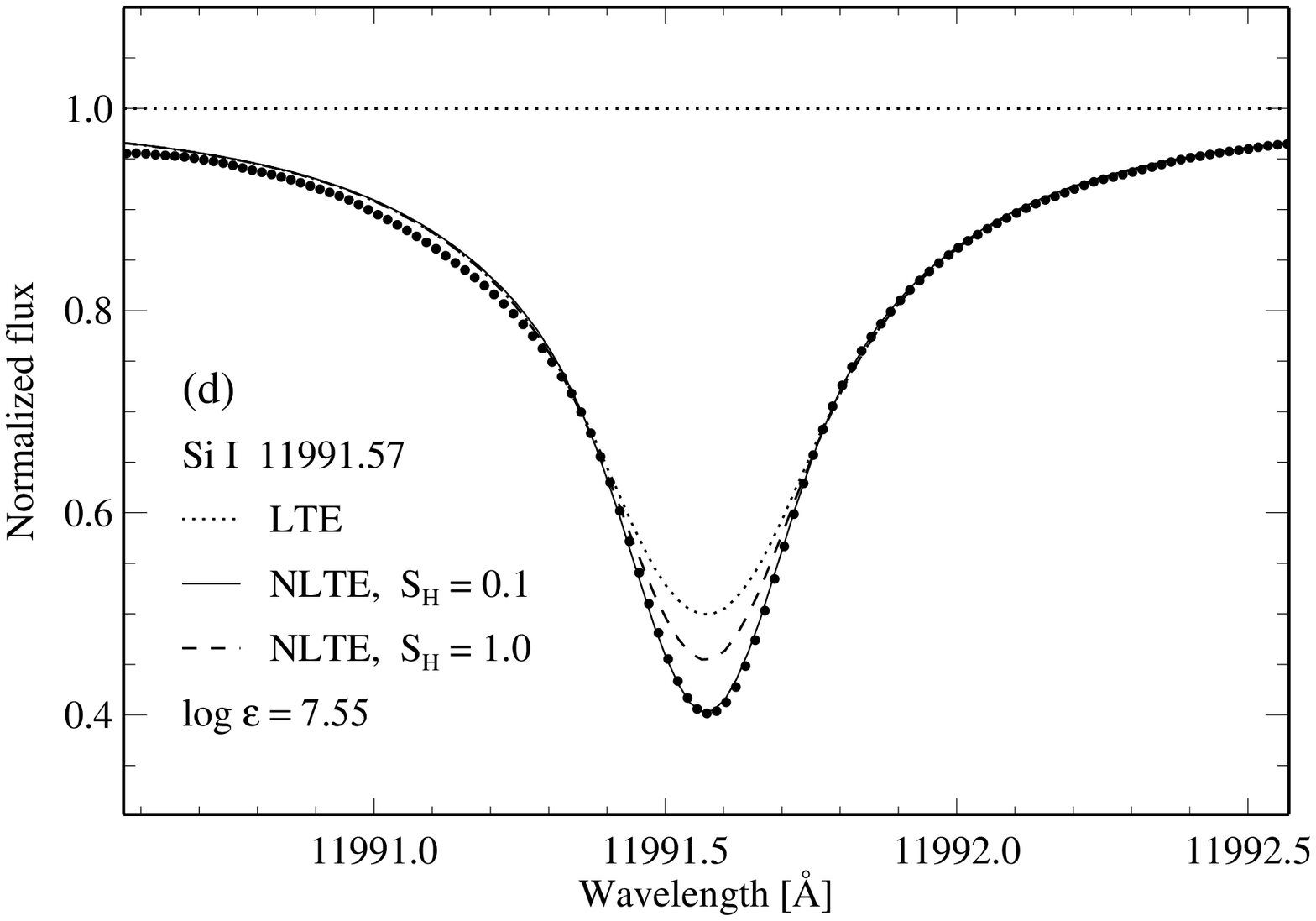}}\\[-0.3cm]
\resizebox{\textwidth}{!}{\includegraphics{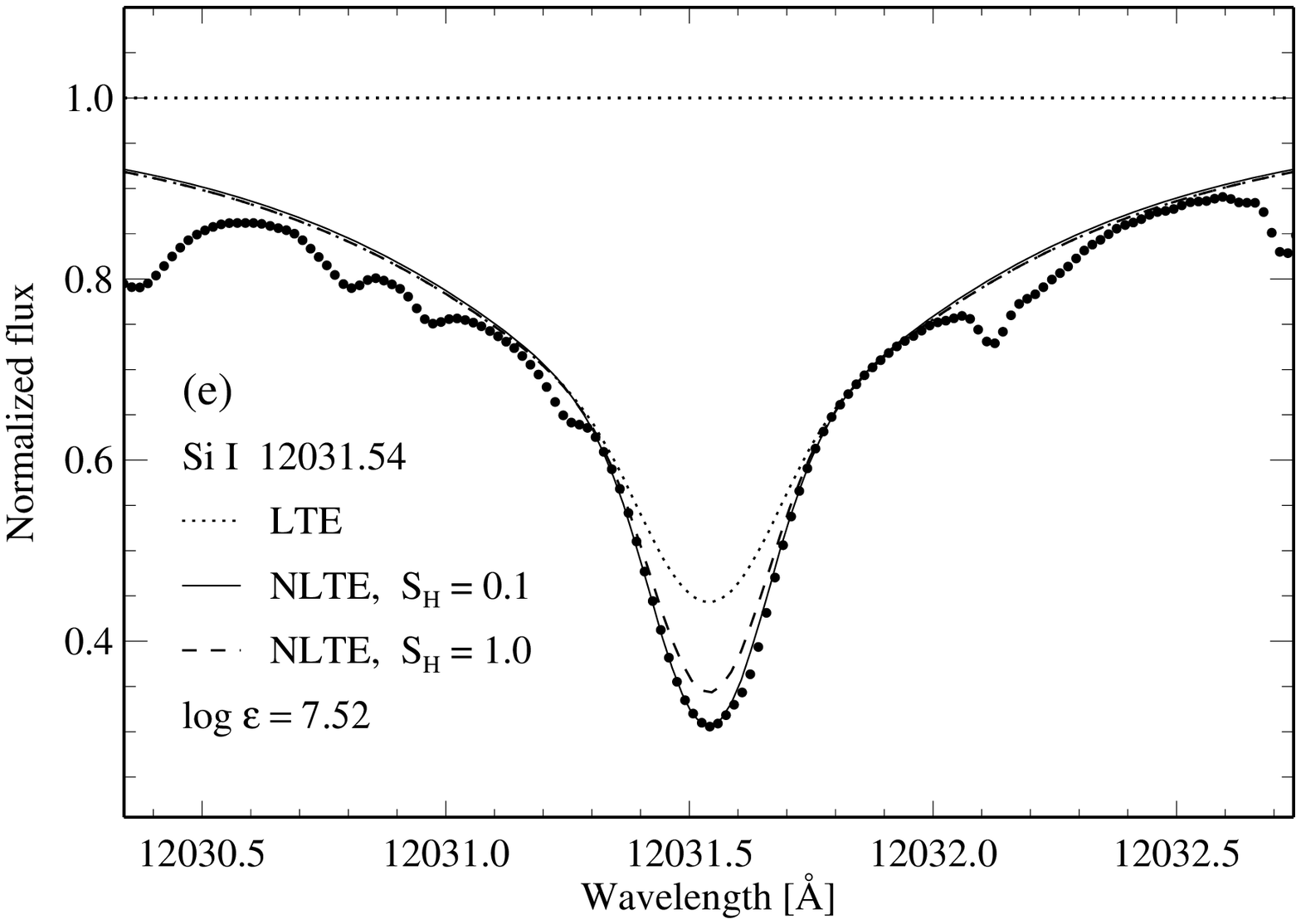}\includegraphics{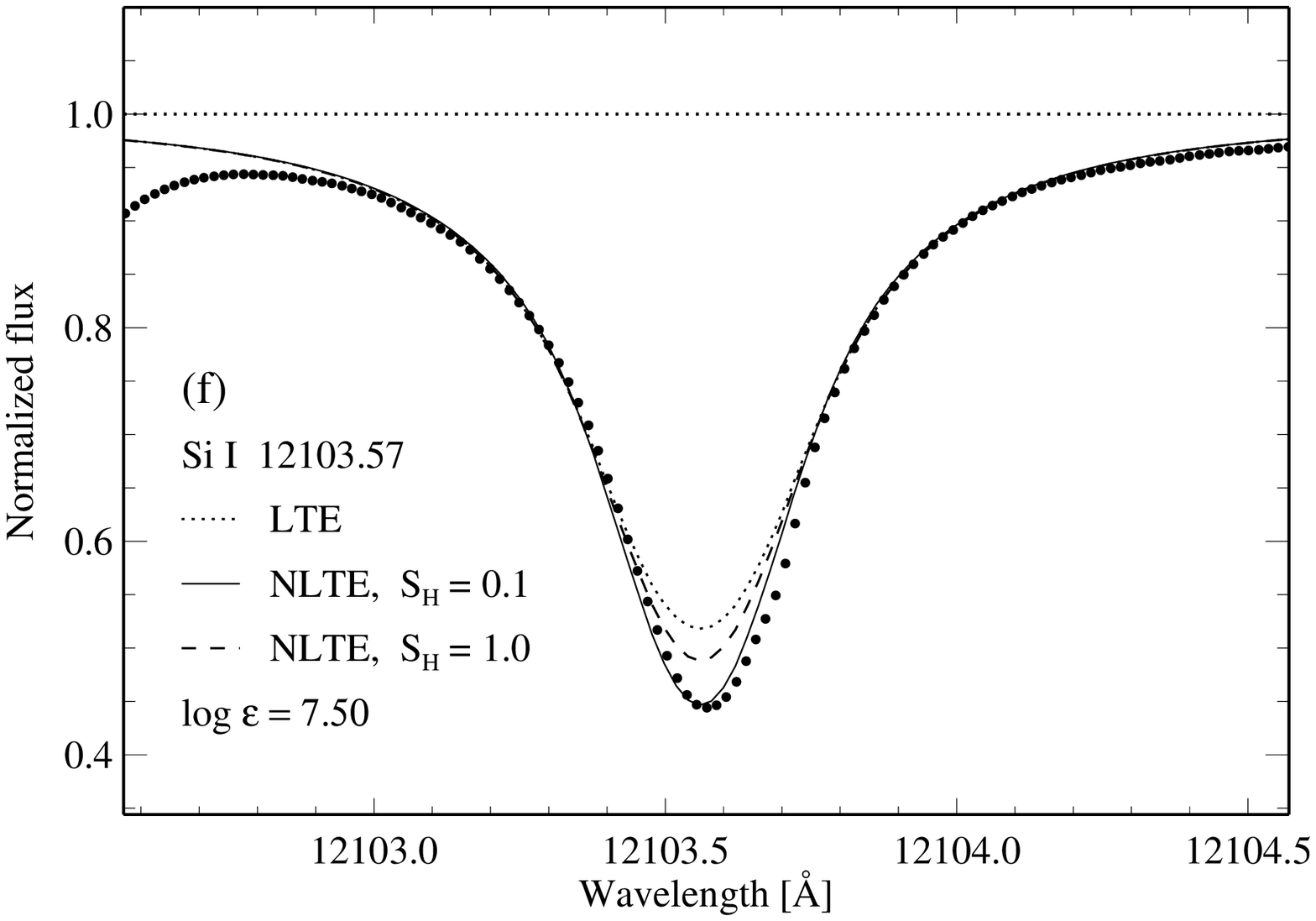}}\\[-0.3cm]
\caption[]{Near-infrared \ion{Si}{i} solar line profiles. NLTE and LTE profiles
are calculated for the same abundance, where the NLTE profile with \SH\ = 0.1
refers to the best fit.} \label{fig6}
\end{figure*}

\section{Line synthesis and the solar Si abundance}

The observed solar flux spectrum was taken from the Kitt Peak Atlas
(Kurucz et al. \cite{K84}). Spectral line synthesis is performed
interactively using the SIU program package of Reetz (\cite{RE91}).
The profiles of all the lines given in Table 1 were calculated for
various assumptions, with atomic parameters varied as described
above. Van der Waals damping constants log $C_6$ for \ion{Si}{i}
lines were computed according to the Anstee \& O'Mara (\cite{AO91},
\cite{AO95}) interpolation tables. For use in other stars the $\log
gf\varepsilon_\odot$ values in the last column are most important.
They are given in Table 1
for our final atomic model statistical equilibrium calculations with
\SH\ = 0.1, and they are used to determine solar Si abundances
below. Van der Waals broadening constants for the \ion{Si}{ii} lines
are purely empirical allowing an optimal fit of the line profiles.
Since the profiles are simultaneously affected by macroturbulence
broadening, similar fits could be obtained also with $\sim 5$ times
smaller damping constants. Due to this bias the resulting abundances
are uncertain by 0.05 dex.

Fig. \ref{fig6} shows some fits of the near-infrared lines between
1.0 and 1.3 $\mu$m. In contrast to the Si lines in the visible, the
cores of the near-infrared lines deviate from the LTE profiles. It
is only this variation with the degree of thermalisation that allows
us to determine a best fit value of \SH\ = 0.1. As mentioned above,
this selection does not affect the visible line profiles.

Nearly all of the Si line profiles observed in the solar spectrum
display some kinematic asymmetry with a red-shifted line core
compared to the fits shown in Fig. \ref{fig6} . This difference does
\emph{not} depend on line blends, for some of which information is
evidently missing. The effect is accompanied in most lines by a
corresponding blue shift of the blue line wings, and sometimes by
observed line \emph{wings} that are both deeper than any profile fit
possible under line synthesis with the currently adopted solar
parameters. We note that a large part of such defects are the result
of our approximate handling of photospheric velocity fields. In a 1D
analysis of spectral line formation, all known atmospheric motions
are replaced by a micro-/macroturbulence approach that can at best
account for a \emph{mean} velocity field. A 3D analysis would
probably remove most of these shortcomings (see Asplund \cite{AS05},
for a review). Modelling the complex granular velocities with a 1D
model, however, would at least require the macroturbulence
velocities $\Xi$ to be adjusted to the regions of line formation,
which are different for weak and strong lines. This is impossible
with a single macroturbulence convolution, because higher $\Xi$
values would be needed for the weak lines than for the strong line
wings.

\begin{figure}
\resizebox{\columnwidth}{!}{\includegraphics{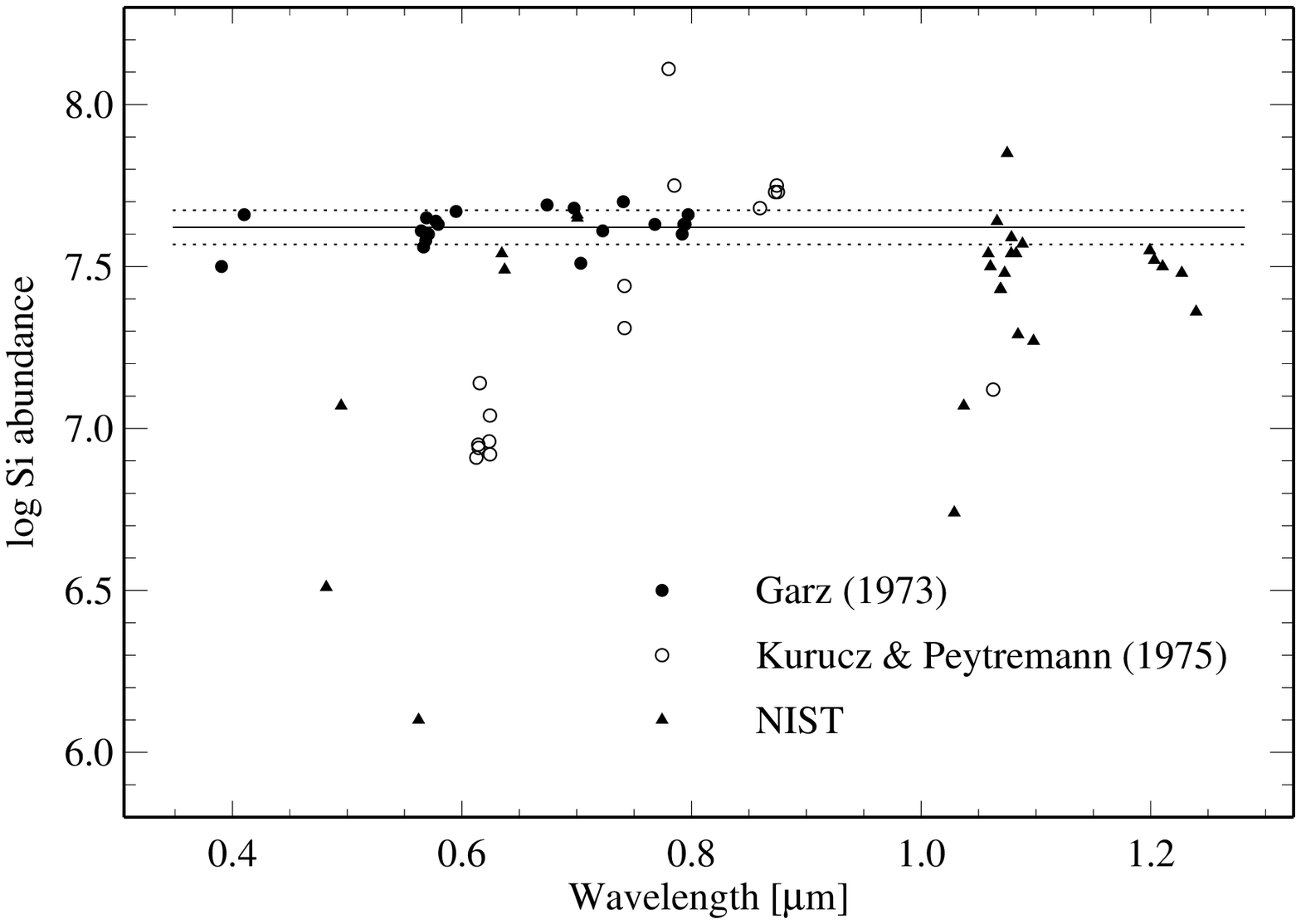}}\\[-3mm]
\resizebox{\columnwidth}{!}{\includegraphics{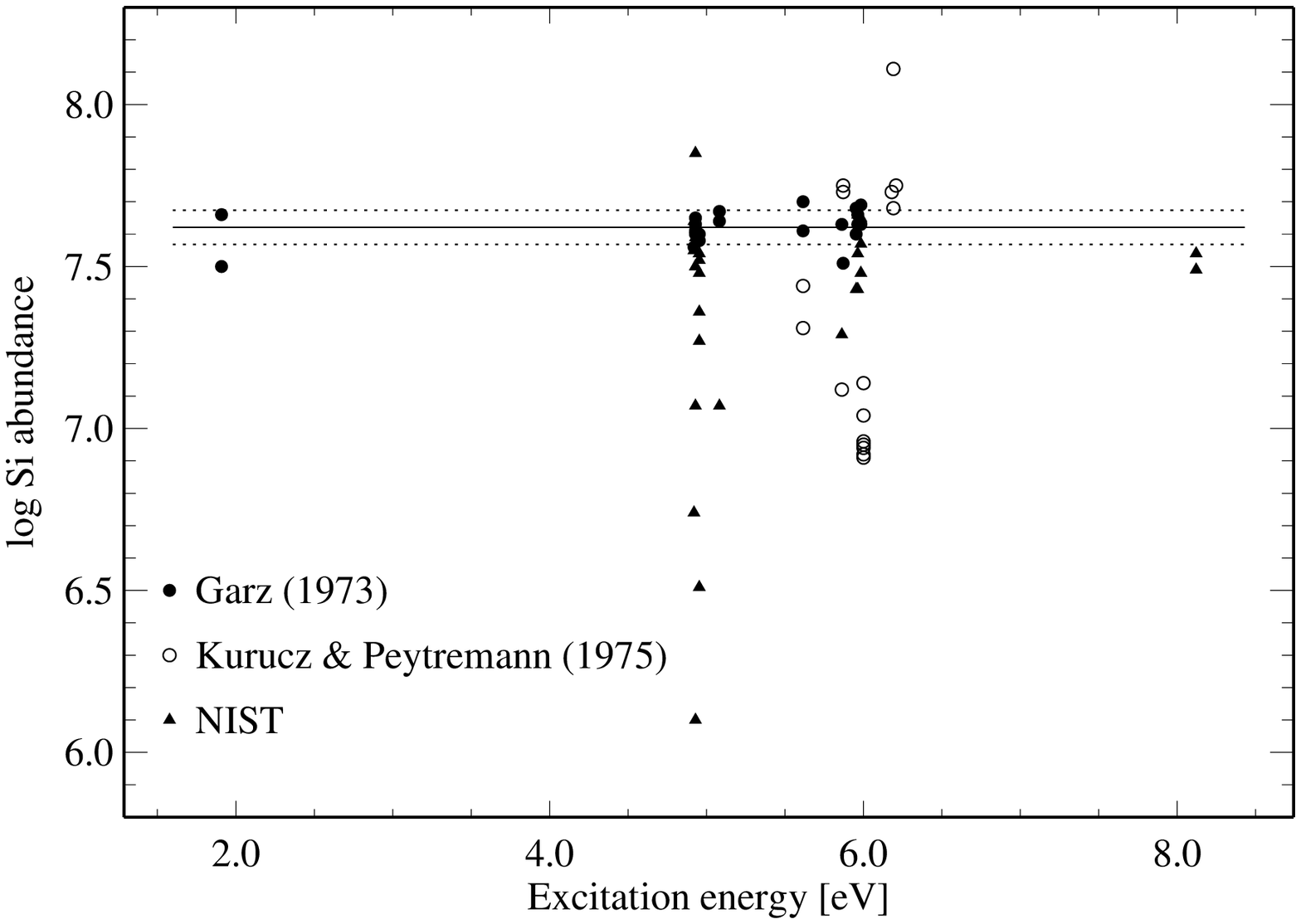}}\\[-3mm]
\resizebox{\columnwidth}{!}{\includegraphics{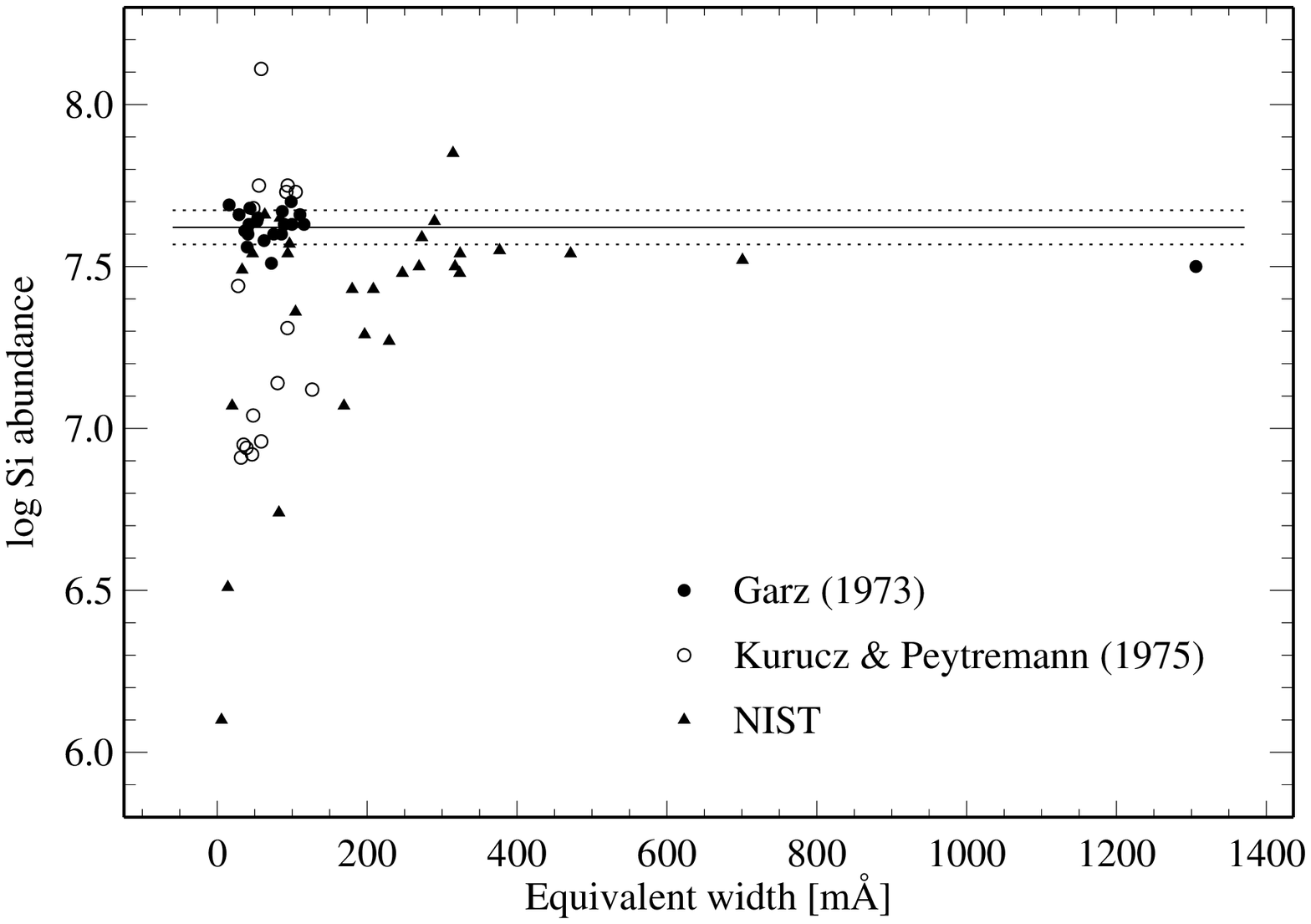}}
\caption[]{Solar abundances calculated with three different sets of
$f$-values. The mean value is shown as a straight line only for the
Garz (1973) data, together with the dotted $\pm 1\sigma$ error
lines} \label{fig7}
\end{figure}

When it comes to spectral line synthesis, every effort to determine
the \emph{absolute} solar photospheric silicon abundance is hampered
by uncertain or inconsistent $f$-values. Since the lifetime
experiments of Savage \& Lawrence (\cite{SL66}) and Marek \& Richter
(\cite{MR73}) there has been an ongoing discussion of the
measurements regarding the lifetime of the \ion{Si}{i}
\Si{4s}{3}{P}{o}{2} level. While all other level lifetimes led to
consistent results, this showed controversial values depending on
the experimental method used (Bashkin et al. \cite{BA80}; Becker et
al. \cite{BE80}; Becker, Zimmermann \& Holweger \cite{BZ80}; Smith
et al. \cite{SM87}). The lifetime measurements of the
\Si{4s}{3}{P}{o}{2} level were carried out with three methods:
\emph{phase shift}, \emph{beam foil}, and \emph{laser fluorescence},
of which the last, used by Becker et al. (\cite{BE80}), should be
the most reliable. However, this value, $\tau = 4.4 \pm 0.4\ {\rm
ns}$ is at variance with the four independent lifetime results
obtained by the other authors, $6.8 \pm 0.8$, $5.95 \pm 0.48$, $5.9
\pm 0.7$ and $6.3 \pm 0.6$ ns, respectively. Thus the
correspondingly calibrated absolute $f$-values differ by roughly
25\%, which translates into a systematic abundance difference of
0.10 dex. This is still the current systematic uncertainty of the
photospheric solar silicon abundance.

To determine the solar Si abundance, it is also necessary to judge
the available $f$-values according to their consistency. In Fig.
\ref{fig7} both the Kurucz and the NIST $f$-values produce an
enormous abundance \emph{scatter} with standard deviations of
$\sigma(\log\varepsilon_\odot) = 0.40$ and $0.38$, respectively. On
the other hand, the relative $f$-values of Garz (\cite{GA73}) are
still the only source for high-quality data of lines in the visible,
with $\log\varepsilon_\odot({\rm Garz}) = 7.62 \pm 0.05$. None of
the sets of oscillator strengths displays a clear trend in either of
the different parameter ranges, although it may be significant that
the three most deviating abundances, based on Kurucz \& Peytremann's
oscillator strengths, emerge from the \Si{4s}{3}{P}{o}{} term, i.e.
the term for which the lifetime measurements were most
contradictory. Removing five outlyers of the NIST abundances from
the sample, the resulting scatter about the mean abundance is
significantly reduced, now with $\log\varepsilon_\odot({\rm NIST}) =
7.52 \pm 0.13$.

\begin{figure}
\resizebox{\columnwidth}{!}{\includegraphics{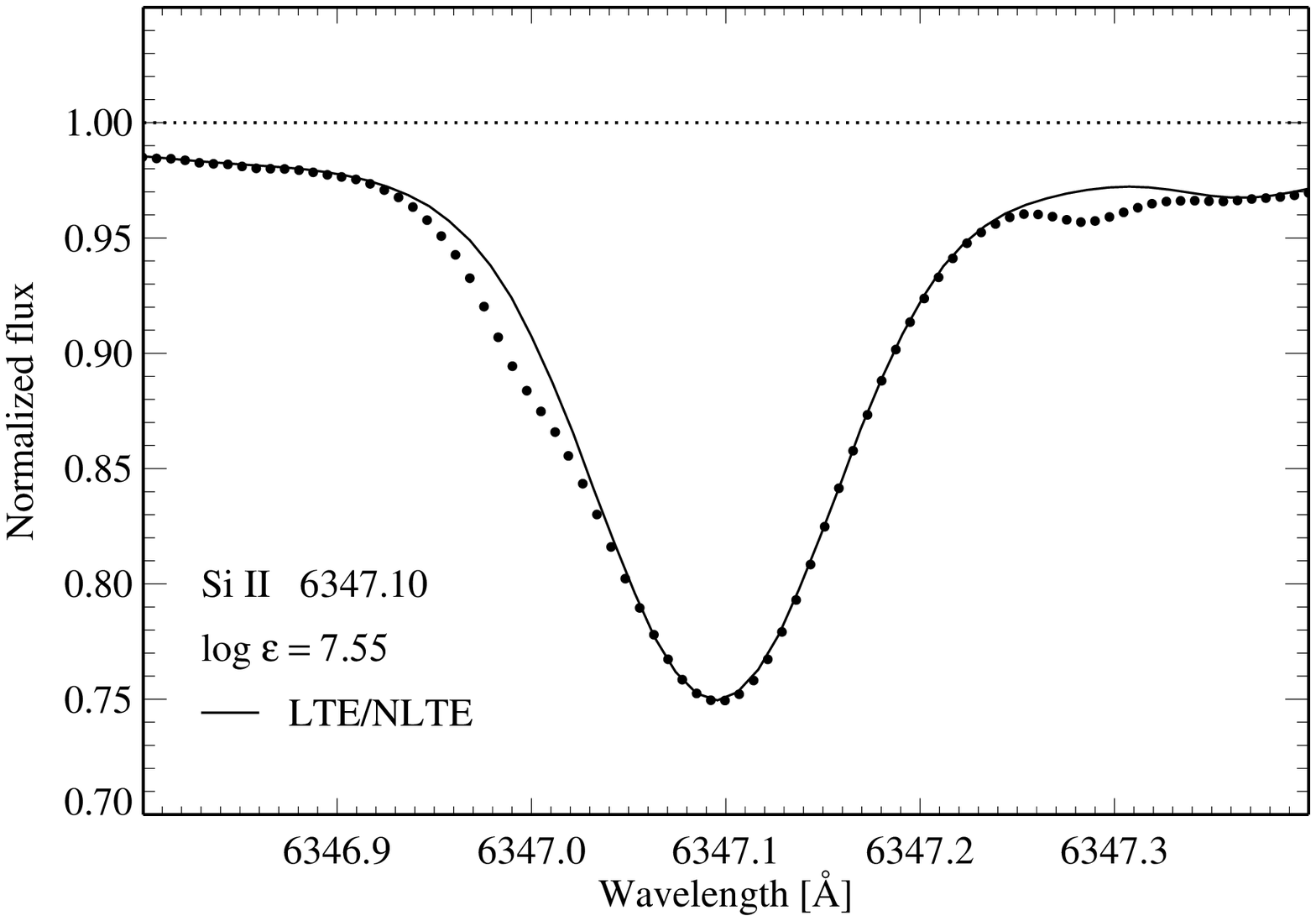}}\\[-3mm]
\resizebox{\columnwidth}{!}{\includegraphics{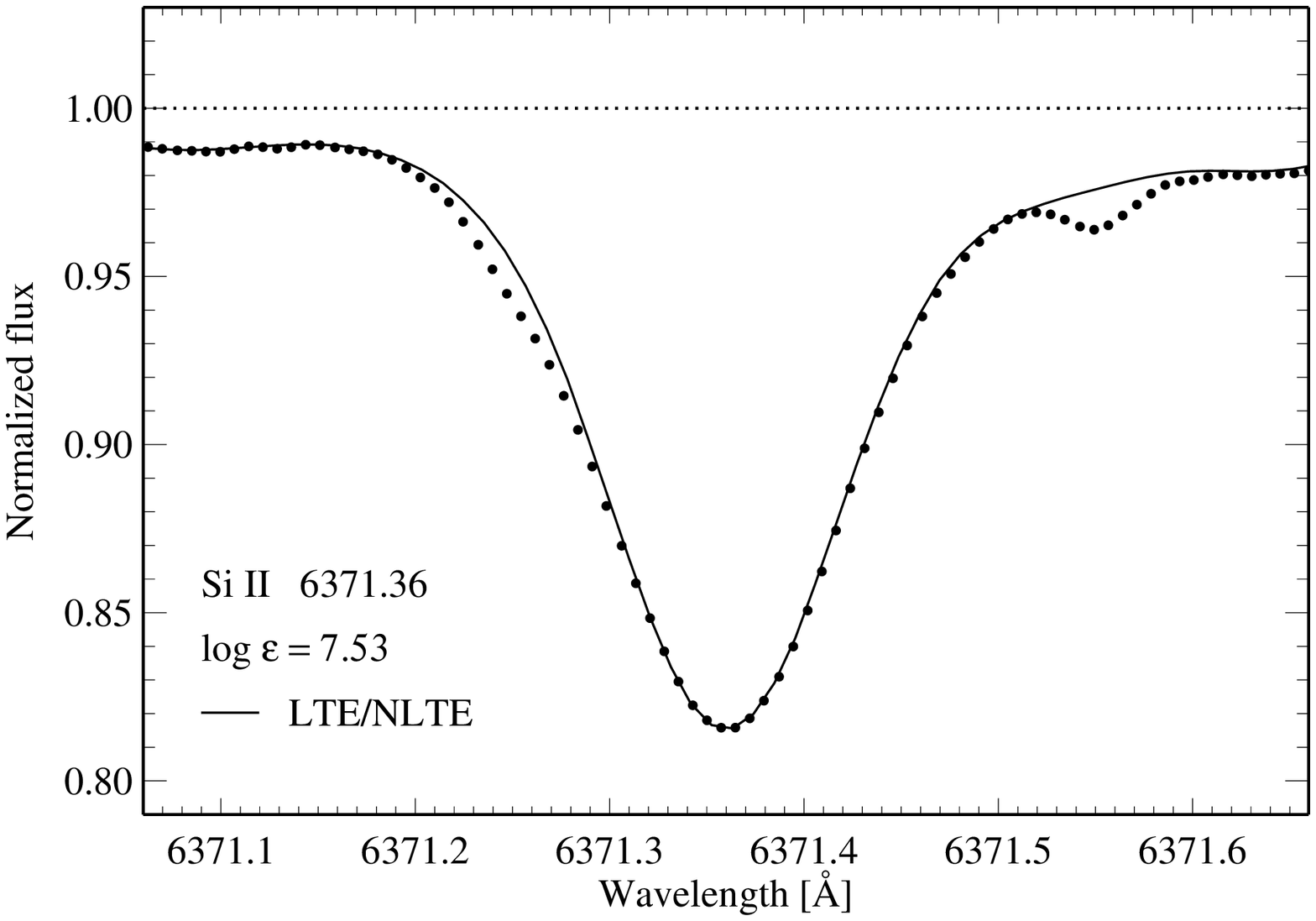}}\\[-3mm]
\caption[]{Solar profile fits for highly excited \ion{Si}{ii} lines at 6347 and
6371 \AA. Blend calculations do not include terrestrial atmospheric lines.}
\label{fig8}
\end{figure}

As seen in Table 1,
it is possible in the solar spectrum to measure the line profiles
absorbed from the \ion{Si}{ii} 4$s^2$S$_{1/2}$ level, which is
excited to more than 8 eV. The results are shown in Fig. \ref{fig8};
the line data and Si abundances are found in Table 1 and Fig.
\ref{fig7}, respectively. Both LTE and NLTE give the same abundances
within 0.01 dex. Although not perfect, the ionisation equilibrium,
i.e. the abundance difference between \ion{Si}{ii} and \ion{Si}{i}
(using Garz' $f$-values) is $\simeq -0.08$ dex, just at the limit of
a $2\sigma$ interval. It is important to note that the two lines
depend sensitively on the choice of the damping constant,
represented here by $C_6$. Using Kurucz' van der Waals broadening
constant which is more than a factor 10 smaller, and replacing the
opacity difference by slightly stronger blends, implies a change in
the abundances by $\simeq 0.05 \ldots 0.10$ dex. Thus the
\ion{Si}{ii} abundances add to the results of the neutral data, but
they are by no means decisive.

Altogether, applying the laser fluorescence lifetime measurements to
the Garz $f$-values would increase the latter by 0.1 dex and thus
reduce the abundance by the same amount. The resulting common mean
abundance together with the NIST values would be
$\log\varepsilon_\odot = 7.52 \pm 0.09$. This would include the
\ion{Si}{ii} lines and lead to a perfect ionisation equilibrium.

\begin{figure}
\resizebox{\columnwidth}{!}{\includegraphics{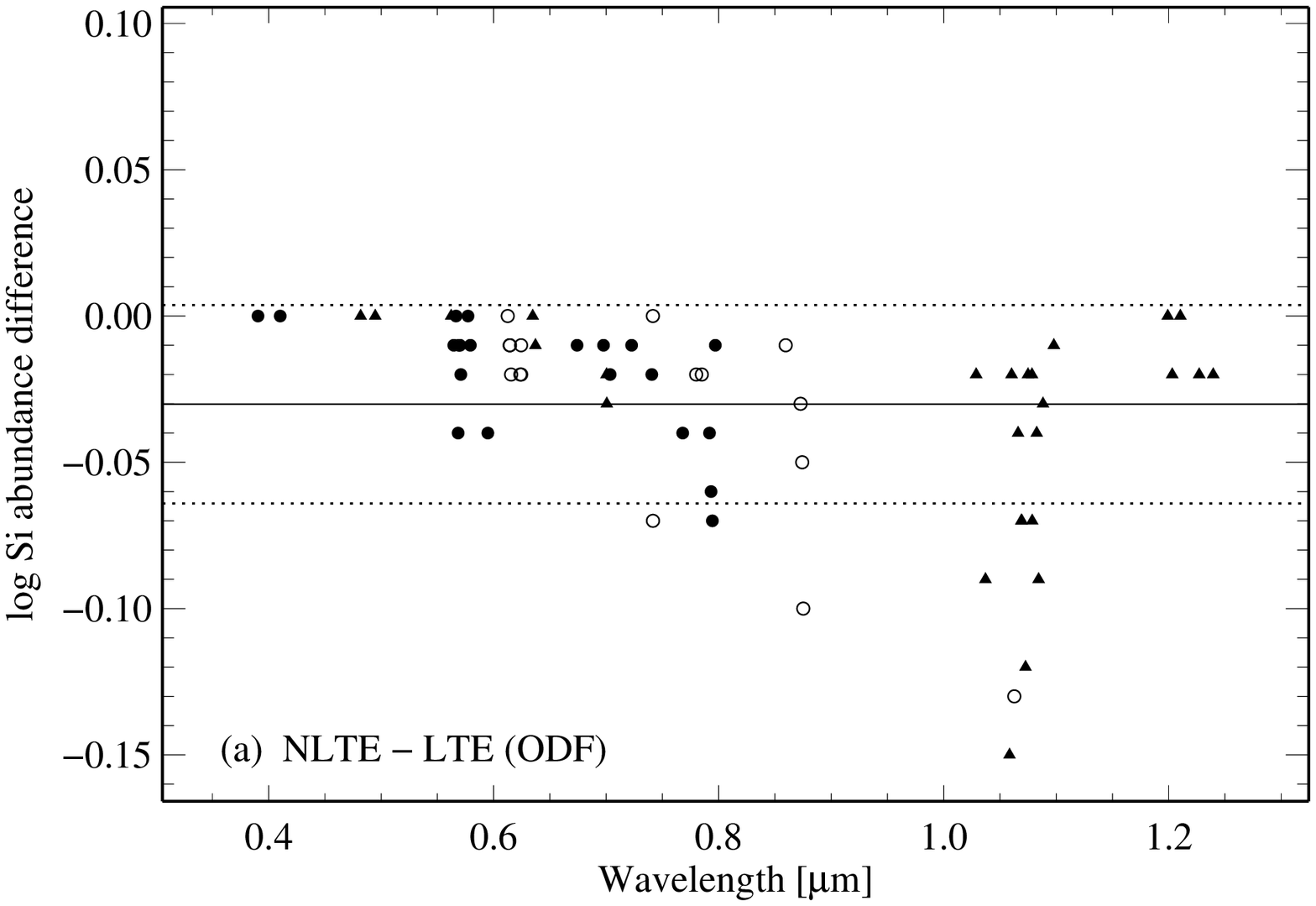}}\\[-3mm]
\resizebox{\columnwidth}{!}{\includegraphics{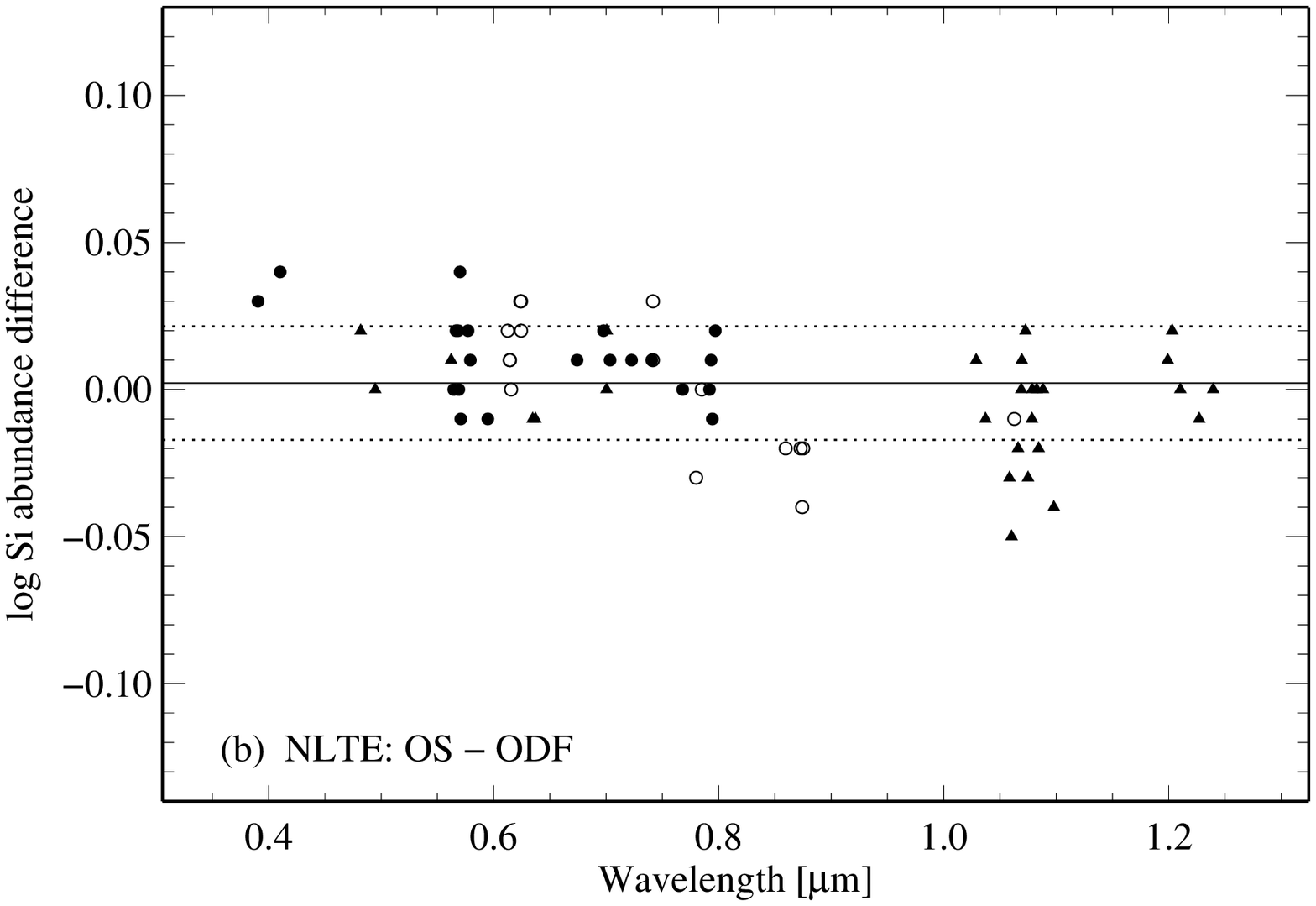}}\\[-3mm]
\caption[]{Logarithmic solar abundance differences for silicon lines.}
\label{fig9}
\end{figure}

The influence of NLTE and of the choice of the model atmosphere on
the solar abundances is best viewed using the line wavelength as an
$x$-axis. Here we present in Fig. \ref{fig9}a the logarithmic
abundance differences between NLTE (final atomic model, \SH\ = 0.1)
and LTE, both from profile fits using the ODF solar model. Although
the mean abundance difference is small, the systematic increase
towards near-infrared wavelengths is remarkable. The lines with low
NLTE abundances all have relatively strong equivalent widths between
$100$ and $350$ m\AA. Their $f$-values are much stronger than those
of the visual lines that arise from the same lower levels, mostly
\Si{4s}{3}{P}{o}{}, \Si{4s}{1}{P}{o}{}, and \Si{4p}{3}{D}{}{},
because the visual lines lead to even higher excited levels (see
Fig. \ref{fig1}). While most of the visible lines are formed well
inside $\log\tau_{5000} = -2$, the region of line formation for the
near-infrared lines is shifted outward to $\log\tau_{5000} = -3
\ldots -2$, where the line source function of those lines differs
significantly from the local Planck function. For the strongest
near-infrared lines, however, the core contributes only marginally
to the abundance determination, which is dominated by the strong
line profile wings coming from a line formation region inside
$\log\tau_{5000} = -2$. Therefore, the trend in Fig. \ref{fig9}a
documents the different line saturation (see also Fig. \ref{fig6}).

Fig. \ref{fig9}b displays the abundance differences calculated for
the final NLTE models between the ODF and the OS solar atmosphere.
Here, we note a similar trend with wavelength, but the corresponding
correlation with the strong lines is much less pronounced. It is
more likely that the different photospheric temperature gradients
(see Fig. \ref{fig3}) contribute to the change in the abundance
differences, because the weaker \ion{Si}{i} lines in the visible are
formed at 50 K higher temperatures and thus require a slightly
higher abundance. The mean Si abundance does not depend on the
choice of the model atmosphere.
\begin{figure}
\resizebox{\columnwidth}{!}{\includegraphics{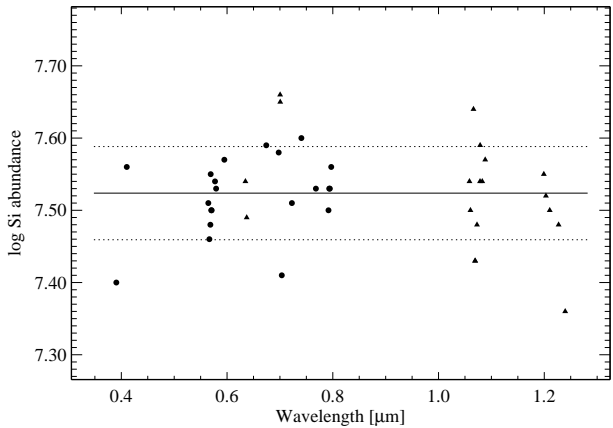}}
\caption[]{Corrected logarithmic solar Si abundances with $f$-values
from Garz (\cite{GA73}, filled circles, corrected by +0.1) or from
NIST (filled triangles, omitting eight outlyers). See text for
discussion.} \label{fig10}
\end{figure}

\section{Discussion}

Both the atomic model and the line formation results do not differ
substantially from the previous analysis of Wedemeyer (\cite{WE01}).
Our aims of atomic model completeness were more directed towards
inclusion of high energy levels, whereas Wedemeyer apparently
favoured level completeness up to a limiting energy. Therefore, he
included a \ion{Si}{i} quintet and some \ion{Si}{ii} quartet terms,
all of which are coupled only by weak forbidden transitions to the
leading multiplets. In fact, our test calculations show that these
terms are insignificant for the departure coefficients.

Therefore it is not too surprising that our results agree in many
details. Most important is our agreement with respect to the nearly
thermal population of all \ion{Si}{i} levels inside optical depth
$\log\tau_{5000} = -2$. In fact, Wedemeyer's departure coefficients
(see his Fig. 2) are even more thermalised than ours. The remaining
differences probably come from slightly different interaction
cross-sections. Our solar silicon abundances are also very similar,
except for an offset that is mostly due to his adjustment of the
Garz (\cite{GA73}) oscillator strengths to the Becker et al.
(\cite{BE80}, \cite{BZ80}) lifetime measurement. We also agree with
his result that in the solar photosphere Si abundance corrections
due to NLTE departures are small. However, it must be noted that
such abundance corrections become significant in the near infrared,
where the cores of strong lines are important.

As is the case for all elements, the agreement between meteoritic
abundances and those obtained from photospheric analysis is a major
topic of research. Because the solar photospheric silicon abundance
is used as a reference for $\alpha$-elements in the Galaxy, and as
the normalising element for meteoritic abundances in the solar
system, it is even more important for the early chemical evolution
of the Milky Way. In particular the question arises as to whether
the meteoritic and photospheric Si/Fe ratios are the same. Fig.
\ref{fig10} combines the selected removal of eight lines with
deviating NIST $f$-values with the correction of the Garz $f$-values
for a 25\% smaller lifetime of the \ion{Si}{i} \Si{4s}{3}{P}{o}{2}
level as measured by Becker et al. (\cite{BE80}). This brings
together the blue, visual and near-infrared wavelength region on a
common abundance scale and avoids excessive scatter. The resulting
photospheric abundance, $\log\varepsilon_\odot({\rm cor}) = 7.52 \pm
0.06$ is in good agreement with the results of Asplund (\cite{AS00})
and Wedemeyer (\cite{WE01}).

During the last years it has become evident that 1D abundance
analyses carry systematic errors as compared with 3D results. While
the 3D results are only a few, and their reliability is not yet
beyond doubt, it seems that some of the metals analysed in 3D or 2D
lead to slightly different photospheric abundances. One of the main
obstacles against a full hydrodynamic investigation is still the
limits of computing speed and memory, respectively. Yet preliminary
calculations cited by Wedemeyer and the 2D simulations of Steffen \&
Holweger (\cite{SH02}) indicate solar abundance corrections for weak
\ion{Si}{i} lines of $\simeq +0.02$, whereas Asplund (\cite{AS00})
favors a 3D hydrodynamical correction of $\simeq -0.02$ for his set
of Si lines, both with reference to the solar model of Holweger \&
M\"uller (\cite{HM74}). The final abundance thus may be between
$7.50$ and $7.54$. With a meteoritic ratio of $\eps{Si}-\eps{Fe} =
0.06$ (Lodders \cite{LO03}; Asplund et al. \cite{AG05}), the
meteoritic Fe abundance would be as low as $7.44 \ldots 7.48$.

\begin{acknowledgements}

This research was supported by the National Natural Science
Foundation of China under grant Nos. 10433010, 10521001, 10778626
and 1071113000052, the National Basic Research Program of China (973
Program) under grant No. 2007CB815103, and by the Deutsche
Forschungsgemeinschaft with grant 446 CHV 112/1,2.

\end{acknowledgements}

\Online
\appendix
\onecolumn
\begin{table}
\caption[]{Atomic data for spectral lines of \ion{Si}{i}
and \ion{Si}{ii} (last two entries). References to the $gf$-values
are G: Garz (\cite{GA73}, experimental), K: Kurucz \& Peytremann
(\cite{KP75}), and N: NIST (http://www.physics.nist.gov/). Values
for $\log gf\varepsilon_\odot$ refer to our final atomic model, with
\SH\ = 0.1.} \label{table1}
\begin{tabular}{lrr@{ -- }lrrr@{~~}lc}
\noalign{\smallskip}\hline\noalign{\smallskip}
 Mult    & $\lambda$~~~~~~  & \multicolumn{2}{c}{Line}   &  $E_{\rm low}$ & $\log C_6$ &  \multicolumn{2}{c}{$\log gf$} & $\log$  \\
         &    [\AA]~~~~   &   \multicolumn{2}{c}{ }      &   [eV]         &            &           &     & $gf\varepsilon_\odot$ \\
\noalign{\smallskip}\hline\noalign{\smallskip}
   2     &  4102.93  &  \Si{3p}{1}{S}{}{0}      &  \Si{4s}{3}{P}{o}{1}  &   1.909    &   --30.972  &  --3.14   & G   &   4.52  \\
   3     &  3905.53  &  \Si{3p}{1}{S}{}{0}      &  \Si{4s}{1}{P}{o}{1}  &   1.909    &   --30.917  &  --1.09   & G   &   6.41  \\
   4     & 12103.56  &  \Si{4s}{3}{P}{o}{1}     &  \Si{4p}{3}{D}{}{1}   &   4.707    &   --30.759  &  --0.35   & N   &   7.15  \\
         & 11991.57  &  \Si{4s}{3}{P}{o}{0}     &  \Si{4p}{3}{D}{}{1}   &   4.920    &   --30.760  &  --0.29   & N   &   7.26  \\
         & 12031.54  &  \Si{4s}{3}{P}{o}{2}     &  \Si{4p}{3}{D}{}{3}   &   4.954    &   --30.743  &    0.41   & N   &   7.93  \\
         & 12270.72  &  \Si{4s}{3}{P}{o}{2}     &  \Si{4p}{3}{D}{}{2}   &   4.954    &   --30.752  &  --0.35   & N   &   7.13  \\
         & 12395.85  &  \Si{4s}{3}{P}{o}{2}     &  \Si{4p}{3}{D}{}{1}   &   4.954    &   --30.757  &  --1.53   & N   &   5.83  \\
   5     & 10603.45  &  \Si{4s}{3}{P}{o}{1}     &  \Si{4p}{3}{P}{}{2}   &   4.707    &   --30.677  &  --0.32   & N   &   7.16  \\
         & 10749.40  &  \Si{4s}{3}{P}{o}{1}     &  \Si{4p}{3}{P}{}{1}   &   4.707    &   --30.689  &  --0.54   & N   &   7.30  \\
         & 10786.88  &  \Si{4s}{3}{P}{o}{1}     &  \Si{4p}{3}{P}{}{0}   &   4.707    &   --30.691  &  --0.42   & N   &   7.16  \\
         & 10661.00  &  \Si{4s}{3}{P}{o}{0}     &  \Si{4p}{3}{P}{}{1}   &   4.920    &   --30.687  &  --0.41   & N   &   7.22  \\
         & 10827.10  &  \Si{4s}{3}{P}{o}{2}     &  \Si{4p}{3}{P}{}{2}   &   4.954    &   --30.677  &    0.17   & N   &   7.71  \\
         & 10979.34  &  \Si{4s}{3}{P}{o}{2}     &  \Si{4p}{3}{P}{}{1}   &   4.954    &   --30.688  &  --0.32   & N   &   6.95  \\
   6     & 10371.30  &  \Si{4s}{3}{P}{o}{1}     &  \Si{4p}{3}{S}{}{1}   &   4.707    &   --30.659  &  --0.41   & N   &   6.65  \\
         & 10288.90  &  \Si{4s}{3}{P}{o}{0}     &  \Si{4p}{3}{S}{}{1}   &   4.920    &   --30.661  &  --0.89   & N   &   5.85  \\
         & 10585.17  &  \Si{4s}{3}{P}{o}{2}     &  \Si{4p}{3}{S}{}{1}   &   4.954    &   --30.659  &  --0.18   & N   &   7.36  \\
   9     &  5793.07  &  \Si{4s}{3}{P}{o}{1}     &  \Si{5p}{3}{D}{}{2}   &   4.707    &   --30.294  &  --2.06   & G   &   5.57  \\
  10     &  5645.61  &  \Si{4s}{3}{P}{o}{1}     &  \Si{5p}{3}{P}{}{2}   &   4.707    &   --30.294  &  --2.14   & G   &   5.47  \\
         &  5690.43  &  \Si{4s}{3}{P}{o}{1}     &  \Si{5p}{3}{P}{}{1}   &   4.707    &   --30.294  &  --1.87   & G   &   5.77  \\
         &  5701.11  &  \Si{4s}{3}{P}{o}{1}     &  \Si{5p}{3}{P}{}{0}   &   4.707    &   --30.294  &  --2.05   & G   &   5.55  \\
         &  5665.55  &  \Si{4s}{3}{P}{o}{0}     &  \Si{5p}{3}{P}{}{1}   &   4.920    &   --30.294  &  --2.04   & G   &   5.51  \\
         &  5708.40  &  \Si{4s}{3}{P}{o}{2}     &  \Si{5p}{3}{P}{}{2}   &   4.954    &   --30.294  &  --1.47   & G   &   6.12  \\
  11     &  5622.23  &  \Si{4s}{3}{P}{o}{1}     &  \Si{5p}{3}{S}{}{1}   &   4.707    &   --30.294  &  --1.64   & N   &   4.46  \\
         &  5684.48  &  \Si{4s}{3}{P}{o}{2}     &  \Si{5p}{3}{S}{}{1}   &   4.954    &   --30.294  &  --1.65   & G   &   5.93  \\
  11.04  &  4818.03  &  \Si{4s}{3}{P}{o}{2}     &  \Si{6p}{3}{D}{}{3}   &   4.954    &   --30.294  &  --1.57   & N   &   4.93  \\
  16     &  5948.54  &  \Si{4s}{1}{P}{o}{1}     &  \Si{5p}{1}{D}{}{2}   &   5.082    &   --30.629  &  --1.23   & G   &   6.43  \\
  17     &  5772.15  &  \Si{4s}{1}{P}{o}{1}     &  \Si{5p}{1}{S}{}{0}   &   5.082    &   --30.287  &  --1.75   & G   &   5.89  \\
  17.09  &  4947.61  &  \Si{4s}{1}{P}{o}{1}     &  \Si{6p}{1}{S}{}{0}   &   5.082    &   --30.287  &  --1.81   & N   &   5.26  \\
  22     &  7415.96  &  \Si{3p^3}{3}{D}{o}{2}   &  \Si{4f}{1}{F}{}{3}   &   5.616    &   --29.969  &  --0.50   & K   &   6.80  \\
  23     &  7405.79  &  \Si{3p^3}{3}{D}{o}{1}   &  \Si{4f}{3}{F}{}{2}   &   5.614    &   --29.869  &  --0.82   & G   &   6.88  \\
  23     &  7415.36  &  \Si{3p^3}{3}{D}{o}{2}   &  \Si{4f}{3}{F}{}{2}   &   5.616    &   --29.869  &  --1.60   & K   &   5.84  \\
  26     &  7226.21  &  \Si{3p^3}{3}{D}{o}{1}   &  \Si{4f}{1}{D}{}{2}   &   5.614    &   --29.800  &  --1.51   & G   &   6.09  \\
  27     &  6244.47  &  \Si{3p^3}{3}{D}{o}{2}   &  \Si{5f}{1}{D}{}{2}   &   5.616    &   --29.868  &  --0.69   & K   &   6.22  \\
  28     &  6237.32  &  \Si{3p^3}{3}{D}{o}{1}   &  \Si{5f}{3}{F}{}{2}   &   5.614    &   --29.869  &  --0.53   & K   &   6.43  \\
  28     &  6243.82  &  \Si{3p^3}{3}{D}{o}{2}   &  \Si{5f}{3}{F}{}{3}   &   5.616    &   --29.868  &  --0.77   & K   &   6.26  \\
  29     &  6145.02  &  \Si{3p^3}{3}{D}{o}{2}   &  \Si{5f}{3}{G}{}{3}   &   5.616    &   --29.869  &  --0.82   & K   &   6.12  \\
  29     &  6155.14  &  \Si{3p^3}{3}{D}{o}{3}   &  \Si{5f}{3}{G}{}{4}   &   5.619    &   --29.869  &  --0.40   & K   &   6.73  \\
  30     &  6125.02  &  \Si{3p^3}{3}{D}{o}{1}   &  \Si{5f}{3}{D}{}{2}   &   5.614    &   --29.869  &  --0.93   & K   &   5.98  \\
  30     &  6142.49  &  \Si{3p^3}{3}{D}{o}{3}   &  \Si{5f}{3}{D}{}{3}   &   5.619    &   --29.869  &  --0.92   & K   &   6.03  \\
  31     & 10843.87  &  \Si{4p}{1}{P}{}{1}      &  \Si{4d}{1}{D}{o}{2}  &   5.863    &   --30.145  &    0.15   & N   &   7.42  \\
  32     & 10627.66  &  \Si{4p}{1}{P}{}{1}      &  \Si{4d}{3}{P}{o}{2}  &   5.863    &   --30.692  &    0.00   & K   &   7.11  \\
  36     &  7680.27  &  \Si{4p}{1}{P}{}{1}      &  \Si{5d}{1}{D}{o}{2}  &   5.863    &   --29.656  &  --0.69   & G   &   6.94  \\
  43     &  8752.01  &  \Si{3d}{1}{D}{o}{2}     &  \Si{4f}{1}{F}{}{3}   &   5.871    &   --30.751  &  --0.52   & K   &   7.19  \\
  44     &  8742.46  &  \Si{3d}{1}{D}{o}{2}     &  \Si{4f}{3}{F}{}{3}   &   5.871    &   --30.916  &  --0.63   & K   &   7.10  \\
  50     &  7034.91  &  \Si{3d}{1}{D}{o}{2}     &  \Si{5f}{3}{G}{}{3}   &   5.871    &   --29.027  &  --0.88   & G   &   6.63  \\
  53     & 10689.73  &  \Si{4p}{3}{D}{}{1}      &  \Si{4d}{3}{F}{o}{2}  &   5.954    &   --29.964  &    0.01   & N   &   7.42  \\
  53     & 10694.27  &  \Si{4p}{3}{D}{}{2}      &  \Si{4d}{3}{F}{o}{3}  &   5.964    &   --29.944  &    0.16   & N   &   7.56  \\
  53     & 10727.43  &  \Si{4p}{3}{D}{}{3}      &  \Si{4d}{3}{F}{o}{4}  &   5.984    &   --29.907  &    0.27   & N   &   7.75  \\
  53     & 10784.57  &  \Si{4p}{3}{D}{}{2}      &  \Si{4d}{3}{F}{o}{2}  &   5.964    &   --29.965  &  --0.72   & N   &   6.81  \\
  53     & 10882.83  &  \Si{4p}{3}{D}{}{3}      &  \Si{4d}{3}{F}{o}{3}  &   5.984    &   --29.945  &  --0.73   & N   &   6.84  \\
  57     &  7918.38  &  \Si{4p}{3}{D}{}{1}      &  \Si{5d}{3}{F}{o}{2}  &   5.954    &   --29.663  &  --0.61   & G   &   6.98  \\
  57     &  7932.35  &  \Si{4p}{3}{D}{}{2}      &  \Si{5d}{3}{F}{o}{3}  &   5.964    &   --29.663  &  --0.47   & G   &   7.15  \\
  57     &  7944.00  &  \Si{4p}{3}{D}{}{3}      &  \Si{5d}{3}{F}{o}{4}  &   5.984    &   --29.663  &  --0.31   & G   &   7.30  \\
  57     &  7970.30  &  \Si{4p}{3}{D}{}{2}      &  \Si{5d}{3}{F}{o}{2}  &   5.964    &   --29.663  &  --1.47   & G   &   6.19  \\
\noalign{\smallskip}\hline\noalign{\smallskip}
\end{tabular}
\end{table}
\begin{table}
{\bf Table 1} (continued)\\
\label{table1} \tabcolsep0.19cm
\begin{tabular}{lrr@{ -- }lrrr@{~~}lc}
\noalign{\smallskip}\hline\noalign{\smallskip}
 Mult    & $\lambda$~~~~~~  & \multicolumn{2}{c}{Line}   &  $E_{\rm low}$ & $\log C_6$ &  \multicolumn{2}{c}{$\log gf$} & $\log$  \\
         &    [\AA]~~~~   &   \multicolumn{2}{c}{ }      &   [eV]         &            &           &     & $gf\varepsilon_\odot$ \\
\noalign{\smallskip}\hline\noalign{\smallskip}
  60     &  6976.51  &  \Si{4p}{3}{D}{}{1}      &  \Si{6d}{3}{F}{o}{2}  &   5.954    &   --29.895  &  --1.17   & G   &   6.50  \\
  60     &  7003.57  &  \Si{4p}{3}{D}{}{2}      &  \Si{6d}{3}{F}{o}{3}  &   5.964    &   --29.434  &  --0.91   & N   &   6.74  \\
  60     &  7005.89  &  \Si{4p}{3}{D}{}{3}      &  \Si{6d}{3}{F}{o}{4}  &   5.984    &   --29.090  &  --0.75   & N   &   6.90  \\
  60.02  &  6741.63  &  \Si{4p}{3}{D}{}{3}      &  \Si{8s}{3}{P}{o}{2}  &   5.984    &   --29.853  &  --1.75   & G   &   5.94  \\
  79     &  8728.01  &  \Si{3d}{3}{F}{o}{2}     &  \Si{5f}{3}{F}{}{3}   &   6.191    &   --29.241  &  --0.61   & K   &   7.12  \\
  80     &  8595.97  &  \Si{3d}{3}{F}{o}{3}     &  \Si{5f}{3}{G}{}{4}   &   6.181    &   --29.125  &  --1.04   & K   &   6.64  \\
  81     &  7800.00  &  \Si{3d}{3}{F}{o}{2}     &  \Si{6f}{3}{F}{}{2}   &   6.191    &   --29.125  &  --1.29   & K   &   6.81  \\
  81     &  7849.97  &  \Si{3d}{3}{F}{o}{3}     &  \Si{6f}{3}{F}{}{2}   &   6.208    &   --29.125  &  --0.96   & K   &   6.79  \\
\noalign{\smallskip}\hline\noalign{\smallskip}
   2     &  6347.10  &  \Si{4s}{2}{S}{}{1/2}    &  \Si{4p}{2}{P}{o}{3/2}&   8.121    &   --30.200  &    0.15   & N   &   7.77  \\
   2     &  6371.36  &  \Si{4s}{2}{S}{}{1/2}    &  \Si{4p}{2}{P}{o}{1/2}&   8.121    &   --30.200  &  --0.08   & N   &   7.45  \\
\noalign{\smallskip}\hline\noalign{\smallskip}
\end{tabular}
\end{table}
\end{document}